\documentclass[final,3p,times]{elsarticle}
\usepackage[latin9]{inputenc}
\usepackage{bm}
\usepackage{amsmath}
\usepackage{amssymb}
\usepackage{graphicx}
\usepackage{esint}
\usepackage{multirow}
\usepackage[subfigure]{graphfig}
\usepackage{rotating}
\usepackage{verbatim}
\usepackage{setspace}
\usepackage{lineno,hyperref}
\usepackage{mathtools,tikz,caption}

\usepackage{float}
\DeclareRobustCommand\sampleline[1]{%
  \tikz\draw[#1] (0,0) (0,\the\dimexpr\fontdimen22\textfont2\relax)
  -- (3em,\the\dimexpr\fontdimen22\textfont2\relax);%
}
\makeatletter

\usepackage{tabularx}\usepackage{subfigure}\usepackage{subfigure}\usepackage{color}
\usepackage{setspace}

\usepackage{bm}

\begin{document}

\title{High-fidelity simulation of regular waves based on multi-moment finite volume formulation and THINC method}

\author[zjuAd]{Zhihang Zhang}

\author[zjuAd]{Xizeng Zhao\corref{mycorrespondingauthor}}
\cortext[mycorrespondingauthor]{Corresponding author.}
\ead{xizengzhao@zju.edu.cn}

\author[sjtuAd]{Bin Xie}

\address[zjuAd]{Ocean College, Zhejiang University, Zhoushan Zhejiang 316021, China}
\address[sjtuAd]{School of Naval Architecture, Department of Ocean and Civil Engineering, Shanghai Jiaotong University, Shanghai, 200240, China}

\begin{abstract}
The performance of interFoam (a widely used solver within OpenFOAM package) in simulating the propagation of water waves has been reported to be sensitive to the temporal and spatial resolution. To facilitate more accurate simulations, a numerical wave tank is built based on a high-order accurate Navier-Stokes model, which employs the VPM (volume-average/point-value multi-moment) scheme as the fluid solver and the THINC/QQ method (THINC method with quadratic surface representation and Gaussian quadrature) for the free-surface capturing. Simulations of regular waves in an intermediate water depth are conducted and the results are assessed via comparing with the analytical solutions. The performance of the present model and interFoam solver in simulating the wave propagation is systematically compared in this work. The results clearly demonstrate that compared with interFoam solver, the present model significantly improves the dissipation properties of the propagating wave, where the waveforms as well as the velocity distribution can be substantially maintained while the waves propagating over long distances even with large time steps and coarse grids. It is also shown that the present model requires much less computation time to reach a given error level in comparison with interFoam solver.
\end{abstract}

\maketitle

\section{Introduction}
The finite volume method (FVM) has gained great popularity in numerical simulations of water waves for its better conservativeness and flexibility to adapt to both structured and unstructured grids. Conventional FVM solvers typically employ second-order accurate spatial discretizations, among which the industry-standard solver STAR-CCM+ and the open-source code interFoam (or other solvers build on interFoam, e.g. waves2Foam~\cite{jacobsen2012wave}, IH-FOAM~\cite{higuera2013realistic} and interFlow~\cite{roenby2017new}) are becoming increasingly popular and have been reported to deal successfully with oceanographic and coastal engineering applications~\cite{jacobsen2014formation, jacobsen2014formation02, brown2016evaluation, higuera2013simulating}. However, wave tanks based on second-order FVM are expected to suffer from numerical dissipation, especially along the direction of wave propagation. Simulations of regular waves using STAR-CCM+ software~\cite{peric2015generation} and interFoam solver~\cite{cha2011numerical, larsen2018performance} demonstrate that the simulated surface elevations are sensitive to the temporal and spatial resolution, decreasing wave heights and phase shifts are observed in those cases with large time steps and coarse grids. The velocity profiles along the vertical cross-section are also examined by some authors~\cite{larsen2018performance, wroniszewski2014benchmarking}, severely overestimated velocities are found near the free surface not only in interFoam simulations but also in the solutions with other Navier-Stokes solvers, including Gerris and Th{\'e}tis.

Refining the grids and time steps is a common strategy to minimize those undesirable effects, commonly hundreds of cells are used within a wavelength~\cite{jacobsen2012wave, larsen2018performance, jin2014numerical, hu2016numerical, windt2017assessment}, whereas refinement will indisputably lead to an increase in computational cost. The cost could be extremely expensive for some cases, for instance, those involving short-period waves, since the relatively higher frequencies cause greater numerical dissipation. In view of this, it is natural to pursue high-order accurate models for the simulations of water waves. Conventional FVM requires wide cell stencil to generate high-order reconstructions, the choice of the stencil, however, is not straightforward on unstructured grids. The numerical schemes with a compact stencil, on the other hand, add degrees of freedoms (DOFs) locally on each cell for high-order reconstructions, and are more flexible and easier to implement on unstructured grids. The representative methods of this sort are the discontinuous Galerkin (DG) method~\cite{cockburn1989tvb}, the spectral finite volume (SV) method~\cite{wang2002spectral} and the CIP (constrained interpolation profile) type multi-moment finite volume methods (CIP/MM FVM)~\cite{ii20054th,xiao2006unified,chen2008shallow}. The VPM scheme~\cite{xie2014multi} is an extension of the CIP/MM FVM to incompressible Navier-Stokes equations on unstructured grids with triangular and tetrahedral elements, it is reported to be a superior trade-off regarding the numerical accuracy and computational cost. In this work, the VPM scheme is utilized to solve the fluid dynamics and the THINC/QQ scheme~\cite{xie2017toward} is used for the free-surface capturing, where the latter is claimed to have comparable solution quality to other existing VOF methods with PLIC geometric interface reconstruction. A numerical framework for incompressible interfacial multiphase flows~\cite{xie2017unstructured} is employed to combine the VPM and THINC/QQ schemes. For wave generation, a mass source function~\cite{lin1999internal} is applied in this work. By positioning the wave-maker inside the solution domain, wave damping~\cite{wei1995time} can be straightfowardly applied to all domain boundaries, wave reflections from the tank walls (or structures of interest in future studies) can thus be eliminated in simulations. Comparisons will be conducted between the present model and interFoam solver in simulating the propagation of regular waves within a two-dimensional (2D) domain.

The rest of this paper is organized as follows. After the description of governing equations and solution procedures for both the present model and interFoam solver in section~\ref{governing}, the computation domain along with the wave generation and absorption methods is demonstrated in section~\ref{setup}. Section~\ref{results} presents the simulation results, followed by a discussion on the numerical schemes in section~\ref{discussion}. This paper is ended with conclusion remarks in section~\ref{conclusion}.

\section{Governing equations and solution methods}
\label{governing}
In this work, the incompressible two-fluid flows with moving interface are solved based on the one-fluid model~\cite{prosperetti2009computational}. The Navier-Stokes equations containing the effects of surface tension and gravity are used for both fluids in the same form,
\begin{equation}
\label{eq:continuity}
\nabla\cdot\mathbf{u}=s(t),
\end{equation}
\begin{equation}
\label{eq:momentum}
\frac{\partial{\rho\mathbf{u}}}{\partial{t}}+\nabla\cdot(\rho\mathbf{u}\otimes\mathbf{u})=-\nabla{p}+(\nabla\cdot(\mu\nabla\mathbf{u})+\nabla\mathbf{u}\cdot\nabla\mu)+\mathbf{F}_{s}-\mathbf{g}\cdot\mathbf{x}\nabla\rho+\mathbf{F}_{d},
\end{equation}
\begin{equation}
\label{eq:vof}
\frac{\partial{\phi}}{\partial{t}}+\nabla\cdot(\mathbf{u}\phi)=\phi\nabla\cdot\mathbf{u}.
\end{equation}
where $\mathbf{u}=(u,v)$ is the velocity vector with components $u$ and $v$ in $x$ and $y$ directions respectively, $p$ is the pressure in excess of the hydrostatic part, $\mathbf{g}$ is the gravity acceleration, $\rho$ the density and $\mu$ the dynamic viscosity coefficient. $\mathbf{F}_{s}$ is the surface tension force formulated by $\mathbf{F}_{s}=\sigma\kappa\nabla\phi$ with $\sigma$ being the surface tension coefficient and $\kappa$ the interface curvature. A mass source term $s(t)$ and a momentum source term $\mathbf{F}_{d}$ have been introduced into the equations for wave generation and absorption respectively, the specific forms of which will be given in section~\ref{setup}. The introduction of the mass source term will be embodied in the computation of Poisson equation and the divergence term in the right hand side of equation~\ref{eq:vof}. A volume-of-fluid (VOF) function is used with an indicator function $\phi(\mathbf{x},t)$ distinguishing two kinds of fluids,
\begin{equation}
\phi(\mathbf{x},t)=
\begin{cases}
1 & \text{water} \\
0 & \text{air} \\
0 < \phi < 1 & \text{interface.}\\
\end{cases}
\end{equation}

In the one-fluid model, it is assumed that the intrinsic fluid properties, such as density and viscosity, are updated based on the VOF function as,
\begin{equation}
\rho=\rho_{1}{\phi}+\rho_{2}(1-\phi)~~\text{and}~~\mu=\mu_{1}{\phi}+\mu_{2}(1-\phi),
\end{equation}
where $\rho_{1}$ and $\rho_{2}$ are the densities, $\mu_{1}$ and $\mu_{2}$ are the dynamic viscosity coefficients of water and air respectively.

\subsection{The present model}
The present model combines two newly developed numerical schemes, namely, the VPM scheme and the THINC/QQ scheme. The fluid dynamic equations are discretized by the multi-moment finite volume method, i.e. VPM scheme, while the VOF transport equation is solved by the THINC/QQ scheme. A numerical framework for interfacial multiphase flows on unstructured grids~\cite{xie2017unstructured} is employed to combine these two schemes. In this framework, the fraction-step method~\cite{chorin1968numerical} is used to update the numerical solution from time level $n$($t=t^{n}$) to $n+1$($t=t^{n}+\Delta{t}$) and the solution procedure is summarized as follows.
\begin{enumerate}
\item Update the velocity field from step $n$($\mathbf{u}^{n}$) to $\mathbf{u}^{*}$ by solving the VOF function~\ref{eq:vof} and the advection term~\ref{eq:advection} simultaneously in the time integration framework of the third-order TVD Runge-Kutta~\cite{gottlieb1998total},
\begin{equation}
\label{eq:advection}
\frac{\partial{(\rho\mathbf{u})}}{\partial{t}}=-\nabla\cdot(\rho\overline{\mathbf{u}}^{n}\otimes\overline{\mathbf{u}}^{n}),
\end{equation}
\item Update the velocity field from $\mathbf{u}^{*}$ to $\mathbf{u}^{**}$ by solving the diffusion terms,
\begin{equation}
\label{eq:diffusion}
\frac{\mathbf{u}^{**}-\mathbf{u}^{*}}{\Delta{t}}=(\nabla\cdot(\mu\nabla\mathbf{u}^{*})+\nabla\mathbf{u}^{*}\cdot\nabla\mu),
\end{equation}
\item Update the velocity field from $\mathbf{u}^{**}$ to $\mathbf{u}^{***}$ by adding the effects of surface tension, gravity force and damping force,
\begin{equation}
\label{eq:external}
\frac{\mathbf{u}^{***}-\mathbf{u}^{**}}{\Delta{t}}=\frac{1}{\rho}(\sigma\kappa\nabla\phi-\mathbf{g}\cdot\mathbf{x}\nabla\rho+\mathbf{F}_{d}),
\end{equation}
\item To make the intermediate velocity field $\mathbf{u}^{***}$ satisfy the mass eqaution~\ref{eq:continuity}, it must be corrected by the following projection step. First, the pressure field at step $n+1$ is obtained by solving Poisson equation,
\begin{equation}
\label{eq:poisson}
\nabla\cdot(\frac{1}{\rho}\nabla{p^{n+1}})=\frac{\nabla\cdot\mathbf{u}^{***}-s(t^{n})}{\Delta{t}},
\end{equation}
then the velocity field is corrected by projecting the pressure field,
\begin{equation}
\label{eq:projection}
\frac{\mathbf{u}^{n+1}-\mathbf{u}^{***}}{\Delta{t}}=-\frac{1}{\rho}(\nabla{p^{n+1}}).
\end{equation} 
\end{enumerate}

\begin{figure}[!htb]
  \centering
    \includegraphics[width=0.7\textwidth]{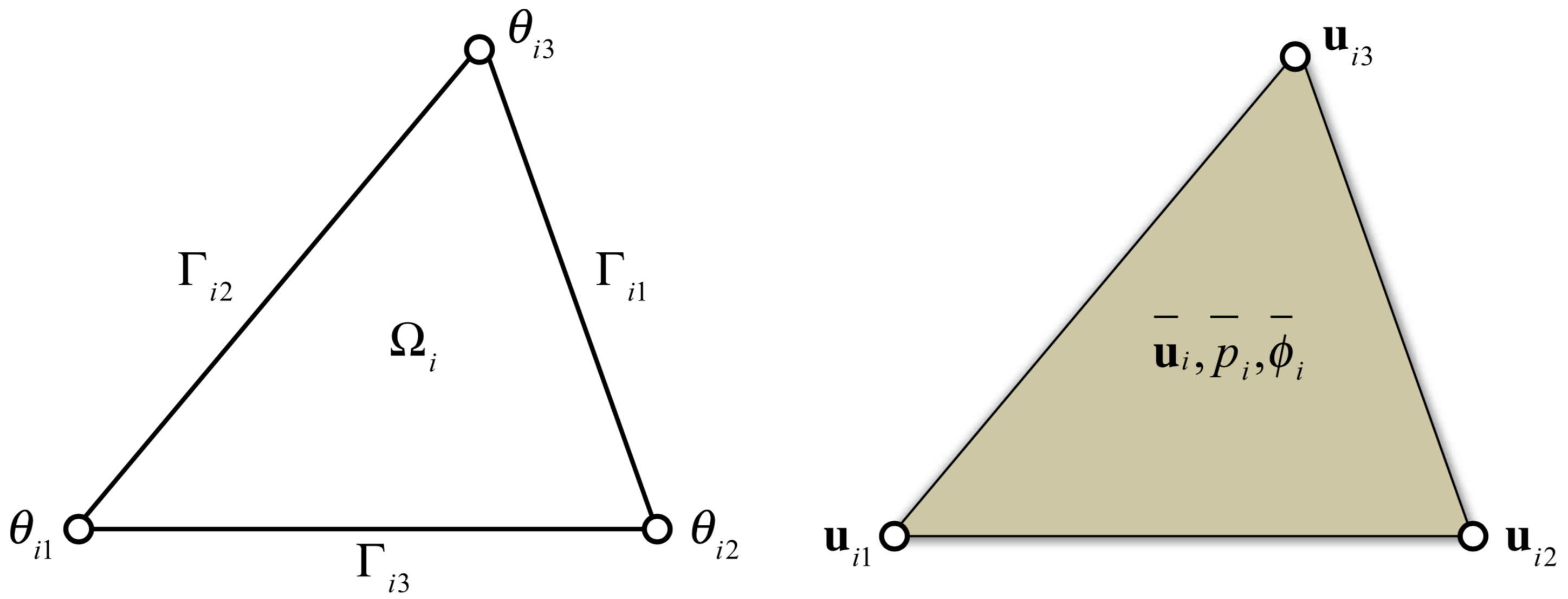}
  \caption{The 2D triangular mesh element (left) and definition of discrete moments (right).}
  \label{fig:MMtriangular}
\end{figure}

The numerical schemes will be explained with a 2D triangular element $\Omega_{i}$ for brevity, more details of the numerical schemes and the 3D formulations can be found in~\cite{xie2014multi, xie2017toward}. Shown in Fig.~\ref{fig:MMtriangular}(left), the vertices are denoted by $\theta_{ik}$ located at $(x_{ik}, y_{ik})~(k=1,2,3)$, the boundary segments are denoted by $\Gamma_{ij}$ where subscript $ij$ denotes the index of the $j$th surface $(j=1,2,3)$.

\subsubsection{VPM scheme for incompressible Navier-Stokes equations}
Different from conventional FVM where only the volume integrated average (VIA) variable is memorized and updated at each time step, the VPM scheme treats both VIA and point value (PV) as the computational variables, such that it realizes a high-order reconstruction with a compact stencil. As shown in Fig.\ref{fig:MMtriangular}(right), in the multi-moment finite volume method, two kinds of moments of the velocity field $\mathbf{u}(x,y,t)$ are used, namely VIA, $\mathbf{\overline{u}}_{i}$ and PVs, $\mathbf{u}_{ik}$ at cell vertices, which are defined by
\begin{equation}
\label{eq:VIAdefinition}
\begin{cases}
\mathbf{\overline{u}}_{i}{(t)}\equiv\frac{1}{|\Omega_{i}|}\int_{\Omega_{i}}{\mathbf{u}(x,y,t)}d{\Omega},\\
\mathbf{u}_{ik}{(t)}\equiv\mathbf{u}(x_{ik},y_{ik},t),~~k=1,2,3;\\
\end{cases}
\end{equation}
where $|\Omega_{i}|$ denotes the volume of cell $\Omega_{i}$. For other physical fields, i.e. pressure field $(\overline{p}_{i})$ and volume fraction $\overline{\phi}_{i}$, only VIA is used as the dicrete moment for simplicity. Given both VIA and PV, a quadratic polynomial for velocity field $\mathbf{u}(x,y)$ on cell $\Omega_{i}$ can be constructed in the local cordinate system as,
\begin{equation}
\label{eq:velocityReconstruction}
U_{i}(\xi,\eta)=\psi_{1}\mathbf{u}_{i1}+\psi_{2}\mathbf{u}_{i2}+\psi_{3}\mathbf{u}_{i3}+\overline{\psi}\overline{\mathbf{u}}_{i}+\psi_{\xi}\mathbf{u}_{\xi{i}}+\psi_{\eta}\mathbf{u}_{\eta{i}},
\end{equation} 
where the unknown coefficients are determined by the corresponding constraint conditions in terms of the multiple moments and are identical for all mesh cells. Both the PVs at the vertices, $\mathbf{u}_{ij}$, and the VIA, $\mathbf{\overline{u}}_{i}$, are updated at every time step through the projection procedure shown above. The derivatives at the cell center, $\mathbf{u}_{\xi{i}}$ and $\mathbf{u}_{\eta{i}}$, can be computed from the PVs and VIAs of the neighboring cells. By constructing high-order polynomial, the accuracy is significantly improved in the computation of the fluid dynamics.

\subsubsection{THINC/QQ scheme for interface capturing}
The discrete form of volume-of-fluid transport equation can be obtained from the finite volume formulation~\ref{eq:vof} over control volume $\Omega_{i}$ as,
\begin{equation}
\label{eq:vofdiscrete}
\frac{\partial\overline{\phi}_{i}(t)}{\partial{t}}+\frac{1}{|\Omega_{i}|}\sum_{j=1}^{J}(v_{n_{ij}}\int_{\Gamma_{ij}}{H(\mathbf{x},t)}_{iup}{d\Gamma})=\frac{\overline{\phi}_{i}(t)}{|\Omega_{i}|}\sum_{j=1}^{J}(v_{n_{ij}}|\Gamma_{ij}|),
\end{equation}
where the volume-of-fluid of each discrete grid element is defined by
\begin{equation}
\label{eq:vofdefinition}
\overline{\phi}_{i}{(t)}=\frac{1}{|\Omega_{i}|}\int_{\Omega_{i}(\mathbf{x})}{H(\mathbf{x},t)}d\mathbf{x},
\end{equation}
and $v_{n_{ij}}=\mathbf{u}_{ij}\cdot\mathbf{n}_{ij}$ denotes the normal velocity on surface $\Gamma_{ij}$, $H(\mathbf{x},t)_{iup}$ stands for the reconstruction function in the upwinding cell of surface $\Gamma_{ij}$. To compute the surface flux $\int_{\Gamma_{ij}}{H(\mathbf{x},t)}_{iup}{d\Gamma}$ in equation~\ref{eq:vofdiscrete}, a hyperbolic tangent function in the local cordinate $(\xi,\eta)$ is used to approximate the indicator function ${H(\mathbf{x},t)}$ for the target cell $\Omega_{i}$ at each time step,
\begin{equation}
\label{eq:THINCfunction}
H_{i}(\xi,\eta)=\frac{1}{2}(1+tanh(\beta(P_{i}(\xi,\eta)+d_{i}))),
\end{equation}
where parameter $\beta$ determines the steepness of the jump in the interpolation function, which has a constant value of 1.5 in this work. $P_{i}(\xi,\eta)=0$ represents the interface surface in the standard element of local element, which is approximated as a curved surface by using fully quadratic polynomial that includes the geometric information,
\begin{equation}
\label{eq:quadratic}
P_{i}(\xi,\eta)=a_{20}\xi^{2}+a_{02}\eta^{2}+a_{11}\xi\eta+a_{10}\xi+a_{01}\eta,
\end{equation}
where coefficients $a_{rs}(r,s=0,1,2~~and~~r+s\leq2)$ are evaluated from the interface normal vector and curvature tensor with least-square method.

The only unknown $d_{i}$ in equation~\ref{eq:THINCfunction} that indicates the location of the interface is determined from VOF values using constrained condition~\ref{eq:vofdefinition},
\begin{equation}
\label{eq:dilocation}
\frac{1}{|\Omega_{i}|}\int_{\Omega_{i}(\mathbf{\xi})}{H(\xi,\eta)}{d}\xi{d}\eta=\overline{\phi}_{i},
\end{equation}
as no general analytical expression is available for integration of multi-dimensional hyperbolic tangent function~\ref{eq:THINCfunction}, a fully multi-dimensional Gaussian quadrature is applied to approximate the integration,
\begin{equation}
\label{eq:quadrature}
\sum_{{g}=1}^{G}\omega_{ig}\bigg(\frac{1}{2}(1+tanh(\beta(P_{i}(\mathbf{\xi}_{ig})+d_{i})))\bigg)=\overline{\phi}_{i},
\end{equation}
where $\mathbf{\xi}_{ig}$ and $\omega_{ig}~~(g=1,2,...,G)$ denote the cordinates and weights of Gaussian points in element $\Omega_{i}$.
To balance the quadrature accuracy and computational cost, the number of quadrature points in each dimension is specified as 3 in this work. Once the indicator function $H_{i}(\xi,\eta)$ is determined, the numerical fluxes on each surface can be calculated by Gaussian quadrature and then used to update the VOF value $\overline\phi_{i}$ by equation~\ref{eq:vofdiscrete}. 

\subsection{interFoam solver}
The solution of the momentum equation in interFoam is performed by constructing a predicted velocity field and then correcting it using the "Pressure Implicit with Splitting of Operators"(PISO) procedures to time advance the pressure and velocity fields. In particular, to evaluate the fluxes on cell boundaries in the computation of advection, a central differencing flux and a non-oscillatory upwind flux are adopted and a TVD conforming flux limiter of $\psi(r)=max[0, min(2r/k,1)]$ is applied to switch between these two flux schemes when using a so-called limitedLinearV scheme. The variable $r$ is an indicator function of consecutive gradients and $k$ is specified as 1 in this work. As central differencing schemes are second-order accurate and upwind schemes are first-order accurate, this solver has its upper limit in solving the advection term as second-order accurate. For interface capturing, the "Multidimensional universal limiter with explicit solution" (MULES) limiter is applied to limit the phase fluxes in solving VOF transport equation~\ref{eq:vof}. And an artificial compression term is added to obtain numerical interface compression, such that it attains the form
\begin{equation}
\label{eq:compressionTerm}
\frac{\partial{\phi}}{\partial{t}}+\nabla\cdot(\mathbf{u}\phi)+\nabla\cdot[\mathbf{u}_{r}\phi(1-\phi)]=\phi\nabla\cdot\mathbf{u}
\end{equation}
The rest of the numerical settings will be those found in the popular dambreak tutorial and interested readers are referred to~\cite{jasak1996error}~\cite{deshpande2012evaluating} for more details.

%
%
\section{Numerical wave tank}
\label{setup}
\begin{figure}[!htb]
  \centering
    \includegraphics[width=0.85\textwidth]{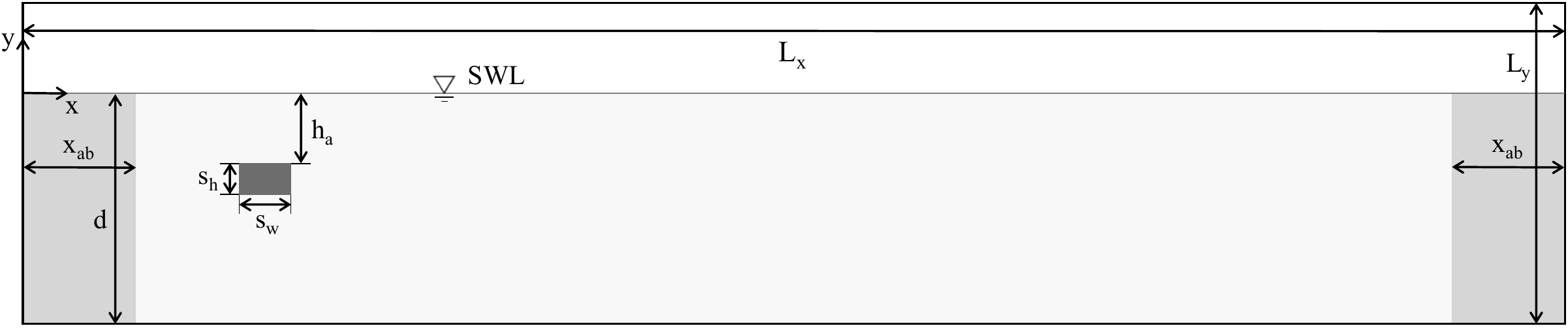}
  \caption{2D computational domain filled with water (ivory white) and air (white), with source region (dark grey) and sponge layers (light grey).}
  \label{fig:domain}
\end{figure}

For this work, a numerical wave tank that has a constant water depth $d$ is built with a rectangular source region located at $x=5$m and an elevation of about $1/3{d}$ from the still water level (SWL), a sponge layer that has a length of $x_{ab}$ is set at each end of the tank. The computational domain along with its coordinate system is presented in Fig.~\ref{fig:domain}. The dimensions of the domain are given in Table~\ref{tab:dimensions}. It is noted that the present wave tank has a relatively large span in horizontal direction (50m), which is designed to facilitate simulations of long-distance wave propagation. 
\vspace{-0.2cm}
\begin{table}[!htb]
  \small
  \caption{Dimensions of the numerical wave tank}
  \centering
    \begin{tabular}{@{\extracolsep{4pt}}lllllll@{}}
    \hline
    \noalign{\smallskip}
    $L_{x}$ & $L_{y}$ & $d$ & $h_{a}$ & $s_{w}$ & $s_{h}$ & $x_{ab}$ \\
    \noalign{\smallskip}
    \hline
    \noalign{\smallskip}
    50.0m  & 0.5m   & 0.35m      & 0.12m& 0.0667m  & 0.04m& 3.0m \\
    \noalign{\smallskip}
    \hline
    \end{tabular}%
  \label{tab:dimensions}%
\end{table}%

As mentioned in section~\ref{governing}, two source functions are embedded in the continuity equation~\ref{eq:continuity} and the momentum equation~\ref{eq:momentum} for wave generation and absorption respectively. The mass source function will be determined by the surface displacement of the target wave~\cite{lin1999internal}, for a linear monochromatic wave with a wave height $H$ and a wave frequency $\sigma$, $\eta(t)={H}sin(\sigma{t})/2$, it will be given as,
\begin{equation}
\label{eq:source}
s(t)=
\begin{cases}
\frac{CH}{A}sin(\sigma{t}) & \text{within source region} \\
0 & \text{elsewhere.} \\
\end{cases}
\end{equation}
where $C$ stands for the phase velocity and $A$ is the area of the source region.

To prevent undesirable wave reflections from the domain boundaries, absorbing regions are employed by adding damping force $\mathbf{F}_{d}$ into the momentum equation, which has a form as
\begin{equation}
\label{eq:damping}
\mathbf{F}_{d}=\rho\mathbf{u}A_{b},
\end{equation}
where $A_{b}$ is the absorbing coefficient, a commonly used form~\cite{wei1995time} is applied in this work,
\begin{equation}
\label{eq:dampingCoeff}
A_{b}=
\begin{cases}
c_{\alpha}\frac{e^{\Big[\big(\frac{|x-x_{st}|}{x_{ab}}\big)^{n_{c}}\Big]}-1}{e^{1}-1} & {x_{st}<x<x_{st}+x_{ab}}\\
0 & \text{elsewhere.}\\
\end{cases}
\end{equation}
where $x_{st}$ and $x_{ab}$ are the starting position and length of the absorbing regions, respectively. $c_{\alpha}$ and $n_{c}$ are the empirical damping coefficients to be determined via the numerical tests, which in this work are specified as $c_{\alpha}=100$ and $n_{c}=3.5$. It is noted that, the steepness parameter $\beta$ from the THINC function is reduced to 0.5 within the absorbing regions to increase the viscosity and consequently to enhance the absorption efficiency. 

\section{Numerical results}
\label{results}
Linear monochromatic waves in a constant water depth are simulated with both the present model and interFoam solver. The quality of simulated waves will be assessed in terms of surface elevations and velocity distribution. The generated waves have a moderate nonlinearity due to the intermediate water depth condition ($kd=1.54$), in this paper, results derived from the second-order Stokes wave theory are referred to as the reference solutions.
\subsection{The base case}
First, a base case is presented, in what follows unless stated the numerical settings will remain in accord with this base case. The parameters of the target wave used to define the mass source function are collected in Table~\ref{tab:linearPara} and a set of grid system discretizes the entire wave tank with roughly 43 cells per wavelength and 10 cells per wave height ($43\times10$) in the vicinity of the SWL. The grid remains uniform in the horizontal direction but nonuniform in the vertical direction, which is coarsened gradually from the SWL to the tank bottom. The time step for the base case is set as $\Delta{t}=0.004s$, which corresponds to 250 time steps per wave period.
\begin{table}[!htb]
  \small
  \caption{Parameters for the linear monochromatic wave used to define the mass source function}
  \centering
    \begin{tabular}{@{\extracolsep{4pt}}lllllll@{}}
     \hline
     \noalign{\smallskip}
     $T_{0}$ & $H_{0}$ & $\lambda_{0}$ & $C_{0}$ & $\sigma_{0}$ & $k_{0}$ & $H_{0}/\lambda_{0}$ \\
     \noalign{\smallskip}
     \hline
     \noalign{\smallskip}
     1.0s  & 0.03m   & 1.425m      & 1.425m/s& 6.2832rad/s  & 4.4092/m& 0.021 \\
     \noalign{\smallskip}
     \hline
    \end{tabular}%
  \label{tab:linearPara}%
\end{table}

\begin{figure}[!htb]
	\begin{minipage}[c]{0.48\linewidth}
		\includegraphics[width=1.0\textwidth]{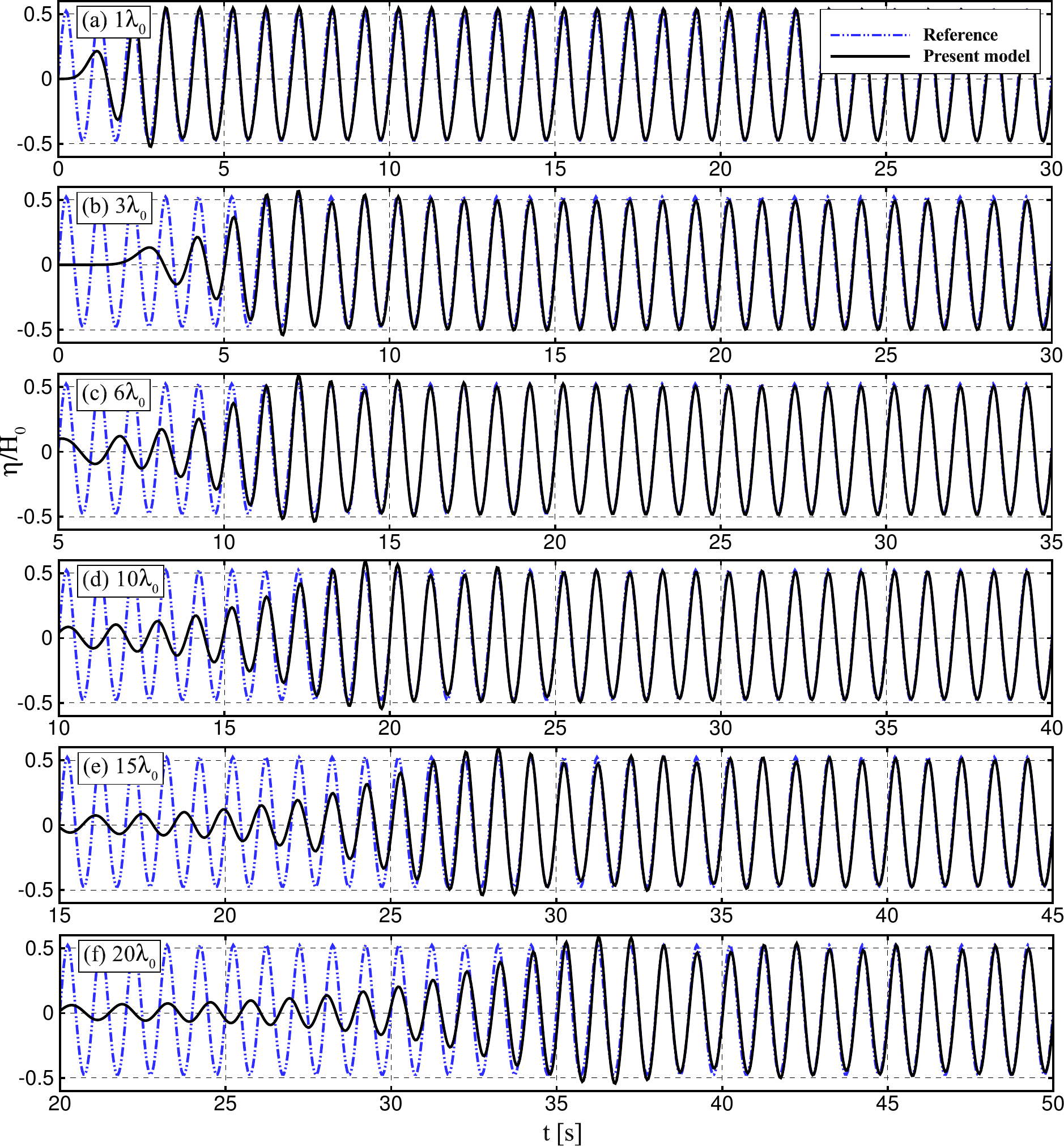}
		\caption{Normalized surface elevations over time at different distances from the source region with the present model.}
	\label{fig:QQVPMgauges}
	\end{minipage}
	\hfill
	\begin{minipage}[c]{0.48\linewidth}
		\includegraphics[width=1.0\textwidth]{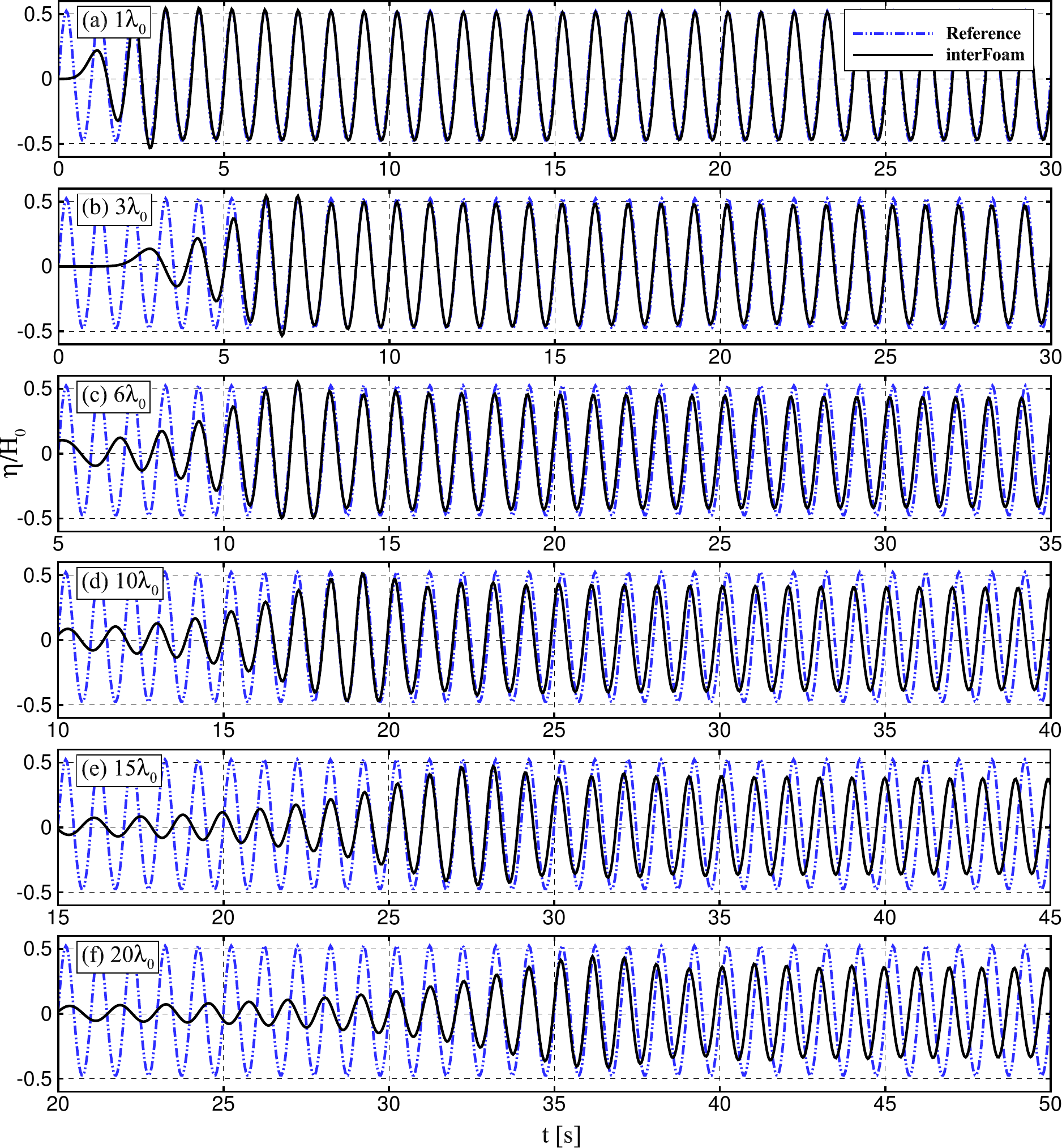}
		\caption{Nomralized surface elevations over time at different distances from the source region with interFoam solver.}
	\label{fig:interFoamgauges}
	\end{minipage}%
\end{figure}

\begin{figure}[!]
	\centering
		\includegraphics[width=0.7\textwidth]{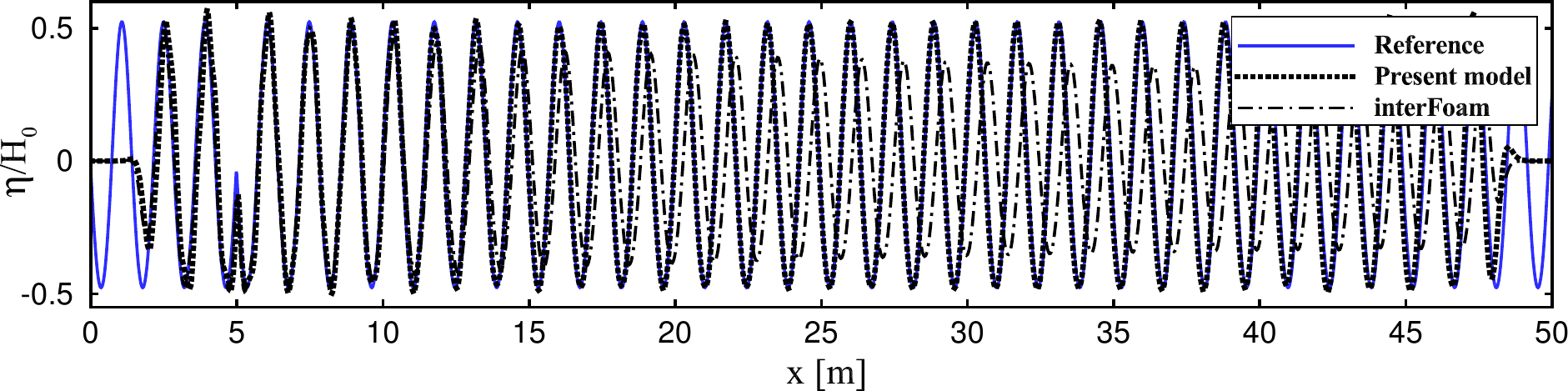}
	\caption{Normalized surface elevations over $x$-location at $t=60T_{0}$.}
	\label{fig:interVSQQAlongx}
\end{figure}

Simulations are performed for 60$T_{0}$ ($T_{0}$ is the nominal wave period). Figs.\ref{fig:QQVPMgauges} and~\ref{fig:interFoamgauges} show the time series of normalized surface elevations at different positions, i.e. $x'=1\lambda_{0}, 3\lambda_{0}, 6\lambda_{0}, 10\lambda_{0}, 15\lambda_{0}, 20\lambda_{0}$, where $x'$ refers to the distance between the center of the source region and the wave gauge, $\lambda_{0}$ is the nominal wavelength. Profiles recorded at $x'=1\lambda_{0}$ (Figs.\ref{fig:QQVPMgauges}a and~\ref{fig:interFoamgauges}a) fit well the reference solution, indicating that both wave makers can produce target waves constantly without significant reflections from the boundaries. However, results from the distant gauges show that the two solvers have utterly different performance in propagating the wave trains. The results with the present model compare well with the reference solutions (Figs.\ref{fig:QQVPMgauges}b-f), the wave shapes are well maintained and no significant phase errors can be observed even after traveling a long distance of $20\lambda_{0}$. While with interFoam solver, generated waves have a noticeable amplitude decay after propagating a distance of $6\lambda_{0}$, in the meantime the simulated profiles start to lead the reference solutions (Figs.\ref{fig:interFoamgauges}b-f). Furthermore, a snapshot of the wave tank at time $t=60T_{0}$ in Fig.\ref{fig:interVSQQAlongx} intuitively shows that with interFoam solver, the wave decay aggravates as the wave trains propagate along the tank and there is also a change in the wavelength, resulting in phase lags over $x$-location. The wave trains simulated with the present model, on the other hand, are well established inside the wave tank, the wave height and wavelength are well maintained and agree well with the reference solutions throughout the working area. Further validation of the wave generation and wave absorption will not be implemented in this work, main attention will be paid to the numerical dissipation in wave propagation.

\subsection{Time step study}
After a preliminary validation of the wave tank, a time step study and a grid study are carried out with the aim of investigating the sensitivity of the solvers to the temporal and spatial resolution. It is noteworthy that in this section, both the time and spatial series of the surface elevations will be plotted to demonstrate the phase consistency with regard to temporal and spatial distribution. Despite the working area of the tank has a length over 20$\lambda_{0}$, the time series recorded at $x'=10\lambda_{0}$ will be assessed in the view of engineering applications. 

\begin{figure}[!]
	\begin{minipage}[t]{0.48\linewidth}
		\centering
		\includegraphics[width=1.0\textwidth]{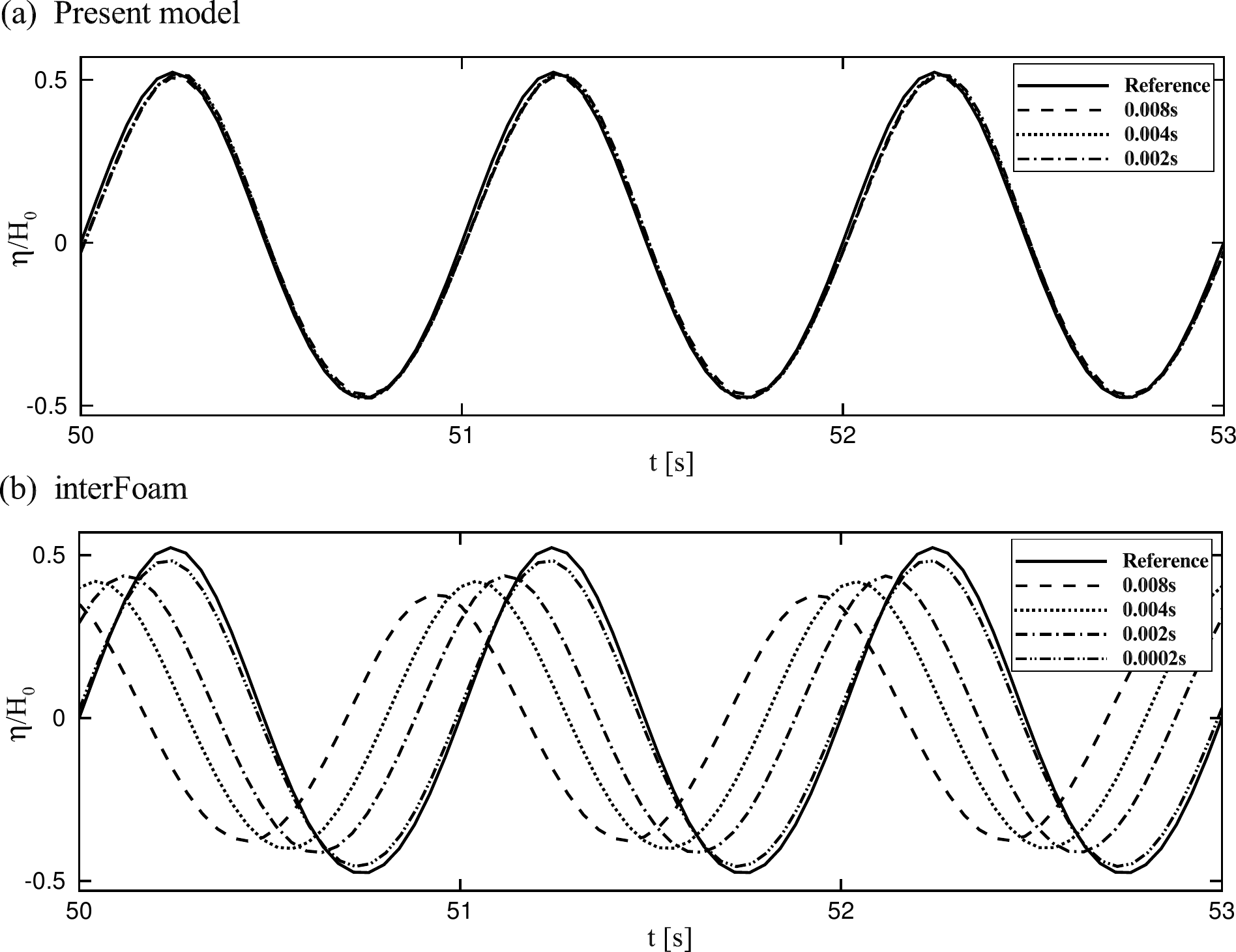}
		\caption{Normalized surface elevations over time recorded $10\lambda_{0}$ away from the source region for different time steps.}
	\label{fig:timestepStudy}
	\end{minipage}
	\hfill
	\begin{minipage}[t]{0.48\linewidth}
		\centering
		\includegraphics[width=1.0\textwidth]{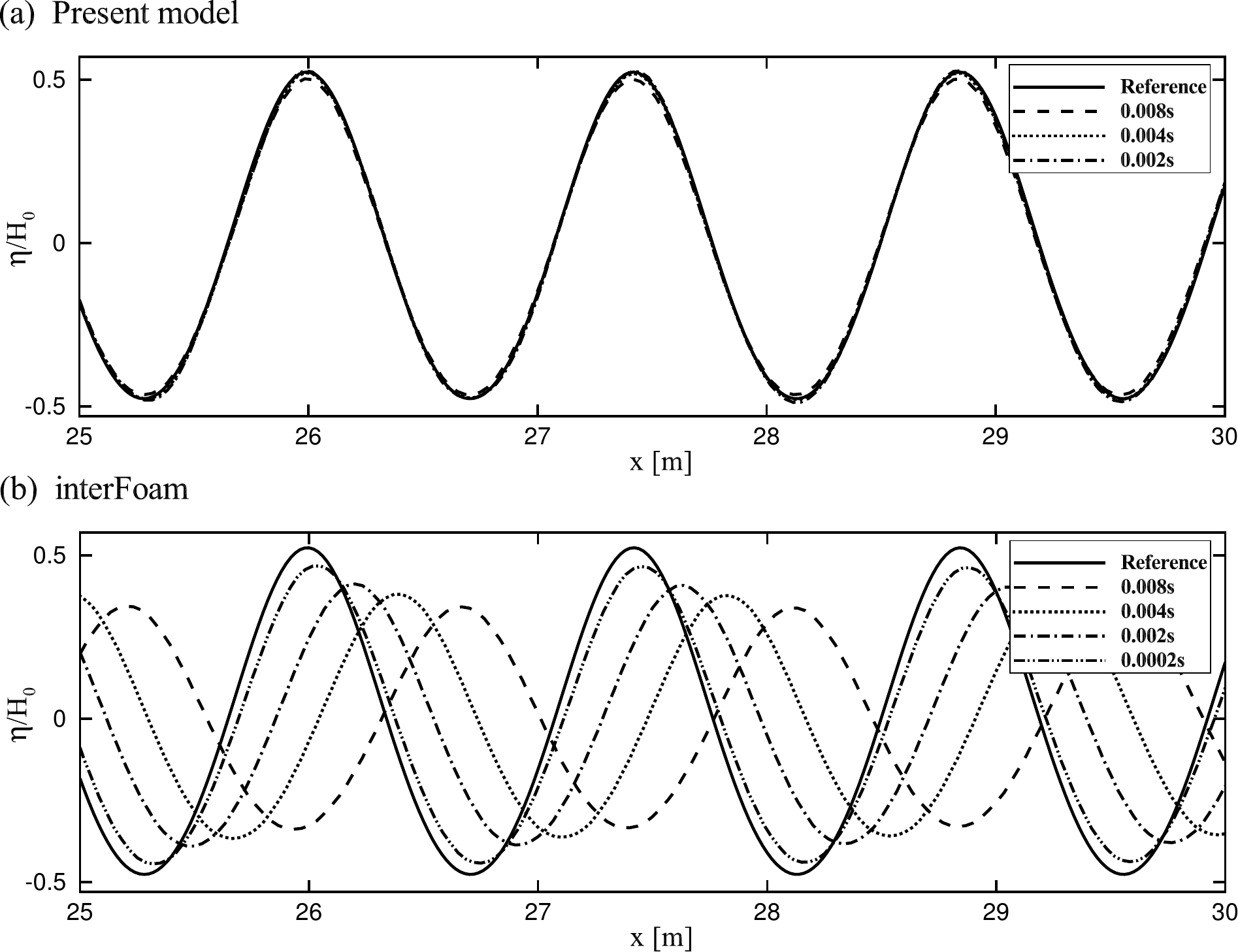}
		\caption{Normalized surface elevations over $x$-location at $t=60T_{0}$ for different time steps.}
	\label{fig:timestepAlongx}
	\end{minipage}
\end{figure}

For the time step study, simulations are performed with different time steps (0.008s, 0.004s and 0.002s) and with otherwise the identical setup to the base case. An enlarged view of the surface elevations recorded at $x'=10\lambda_{0}$ is presented in Fig.\ref{fig:timestepStudy} and a snapshot of the wave tank at $t=60T_{0}$ is shown in Fig.\ref{fig:timestepAlongx}. It is demonstrated that the simulations with the present model are basically insensitive to the temporal resolution, the results are substantially consistent with the reference solutions even using a large time step of 0.008s. The interFoam simulations, on the other hand, show significant wave decay and phase shifts in the 0.008s case and the wave profiles approach the reference solutions as the time step reduces, but noticeable deviation exits even with an extremely small time step of 0.0002s that roughly corresponds to a maximum Courant number of 0.01. 


The phase errors in the propagation of regular waves simulated with interFoam solver have also been reported in~\cite{larsen2018performance} and~\cite{paulsen2014forcing}, they are supposed to arise from the changes in wave periods and wavelengths, so we collect the normalized wave period $T/T_{0}$ and the normalized wavelength $\lambda/\lambda_{0}$ in Table~\ref{tab:parameter}. The simulated wave period $T$ is computed by averaging that of the last 30 waves recorded at $x'=10\lambda_{0}$, while the predicted $\lambda$ is calculated from the surface elevations along the wave tank at $t=60T_{0}$. It can be seen from the table that the present model yields a constant wave period and wavelength that are highly consistent with the nominal values. While reduced wave periods and expanded wavelengths are found in the interFoam simulations, the disparities increase with the time step and have the maximum values in the 0.008s case, including a decrease of 0.4$\%$ in wave periods and a wavelength increase of 2.5$\%$. These changes in wave periods and wavelengths account for the phase shifts found in the surface elevations, though the changes are not significant, noticeable phase shifts are produced due to the cumulative effects. The normalized wave height $H/H_{0}$ is collected in the table as well, where the strategy for evaluating $T$ is utilized again to compute the simulated wave height $H$. It can be seen that the wave decay rates reach the maximum values when a large time step of 0.008s is applied, which are 2.6$\%$ and 26.8$\%$ for the present model and interFoam solver, respectively.
\begin{table}[!htb]
  \small
  \caption{Parameters of simulated waves for the time step study, normalized by the nominal values}
  \centering
    \begin{tabular}{@{\extracolsep{4pt}}llllllll@{}}
    \hline
     \noalign{\smallskip}
     & \multicolumn{3}{l}{Present model} & \multicolumn{4}{l}{interFoam solver}\\
     \noalign{\smallskip}
     \cline{2-4}\cline{5-8}
     \noalign{\smallskip}
     Resolution & 0.008s & 0.004s & 0.002s & 0.008s & 0.004s & 0.002s & 0.0002s \\
     \noalign{\smallskip}
     \hline
     \noalign{\smallskip}
     $T/T_{0}$  & 1.000  & 1.000  & 1.000  & 0.996  & 0.996  & 0.997  & 1.000  \\
     \noalign{\smallskip}
     $\lambda/\lambda_{0}$ & 0.999  & 0.999  & 0.999 & 1.025  & 1.013  & 1.005  & 1.000  \\
     \noalign{\smallskip}
     $H/H_{0}$  & 0.974  & 0.997  & 0.997  & 0.732  & 0.801  & 0.849  & 0.938  \\
     \noalign{\smallskip}
     \hline
    \end{tabular}%
  \label{tab:parameter}%
\end{table}

\begin{figure}[H]
	\centering
		\includegraphics[width=0.5\textwidth]{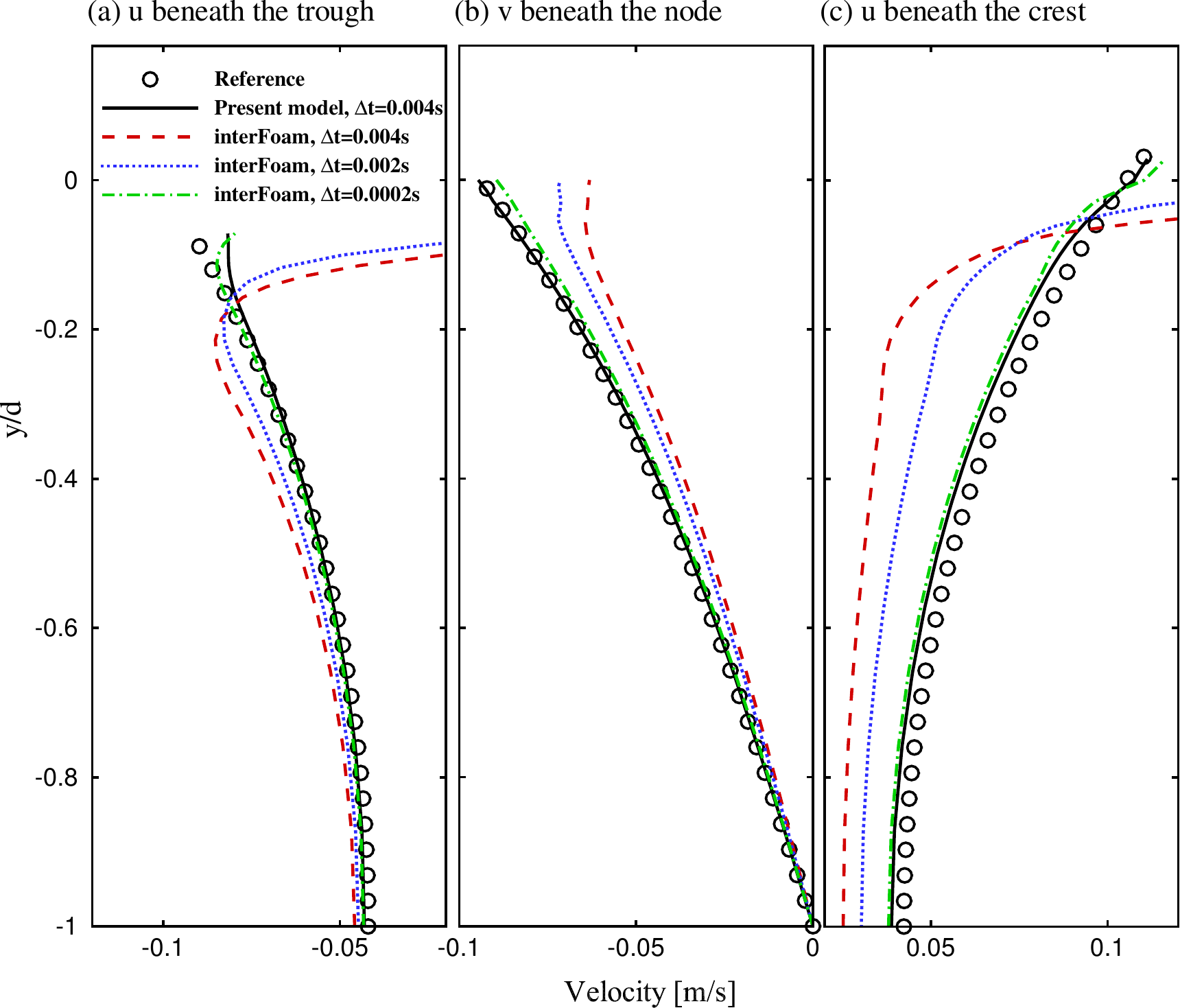}
		\caption{Vertical distribution of velocity components for different time steps with both solvers, recorded at different positions of the propagating wave, at $t=60T_{0}$.}
	\label{fig:UVtimestep}
\end{figure}

Also of great interest is the velocity distribution beneath the free surface, as the water particle kinematics provide the basis for force calculations on coastal and offshore structures, while also relating to e.g. shear stresses and hence sediment transport predictions. In Fig.\ref{fig:UVtimestep}, the velocity profiles along the vertical cross-section are plotted for three characteristic positions, namely, the trough, the node and the crest of the progressive waves. Reference solutions derived from the potential function for a second-order Stokes wave are shown together for comparison. The velocity profiles for the present model using a time step of 0.004s compare well with the reference solution, only slightly underestimated velocities can be found beneath the wave crest in Fig.\ref{fig:UVtimestep}c. While the interFoam simulations display different features. In the 0.004s case, the vertical velocity $v$ beneath the wave node is underestimated (Fig.\ref{fig:UVtimestep}b), meanwhile the horizontal velocity $u$ beneath the wave trough and the wave crest is underestimated in the bulk of the water, especially in the vicinity of the free surface, is severely overestimated (Figs.\ref{fig:UVtimestep}a and c). By reducing the time step to 0.0002s, interFoam solver could have comparative solutions with the present model. The presence of spurious currents near the free surface has been reported by other authors in the cases of both surface tension driven flows and gravity driven flows~\cite{roenby2017new, larsen2018performance, francois2006balanced, wemmenhove2015numerical}. Especially, in a solitary wave simulation by Wroniszewski et al.~\cite{wroniszewski2014benchmarking}, it is shown that the spurious currents around the free surface were developed not only with OpenFOAM code but also in the cases of Gerris and Th{\'e}tis. The reason for such behavior is believed to arise from an imbalance in the discretized momentum equation near the free surface and may explain the decreasing wave periods and increasing wavelengths in the interFoam simulation. The exact mechanism, however, is not yet clear, and will be pursued in further study.

\subsection{Grid study}
\label{GridStudy}
\begin{figure}[!]
	\begin{minipage}[t]{0.48\linewidth}
	\centering
		\includegraphics[width=1.0\textwidth]{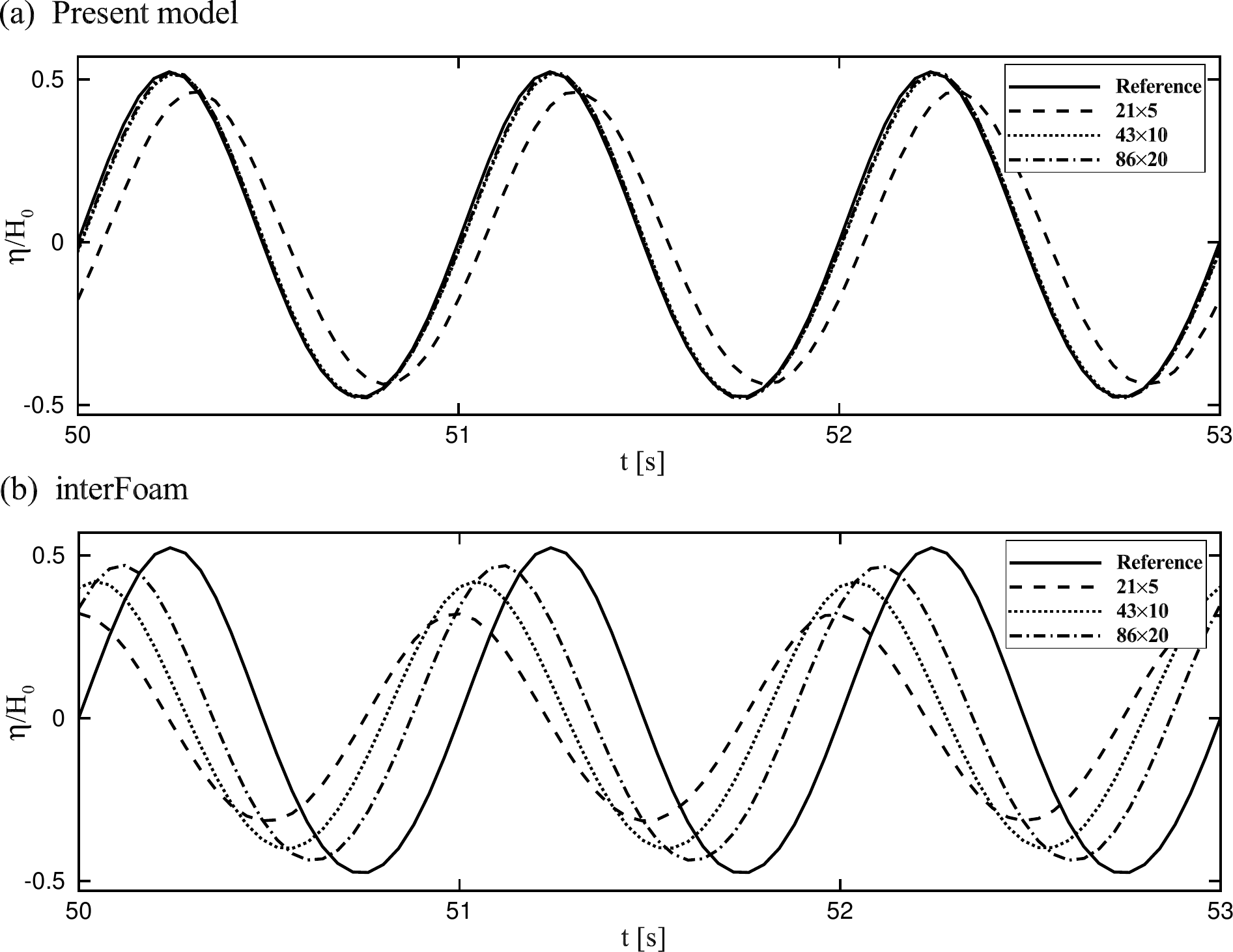}
		\caption{Normalized surface elevations over time recorded $10\lambda_{0}$ away from the source region for different grid resolution.}
	\label{fig:gridStudy}
	\end{minipage}
	\hfill
	\begin{minipage}[t]{0.48\linewidth}
	\centering
		\includegraphics[width=1.0\textwidth]{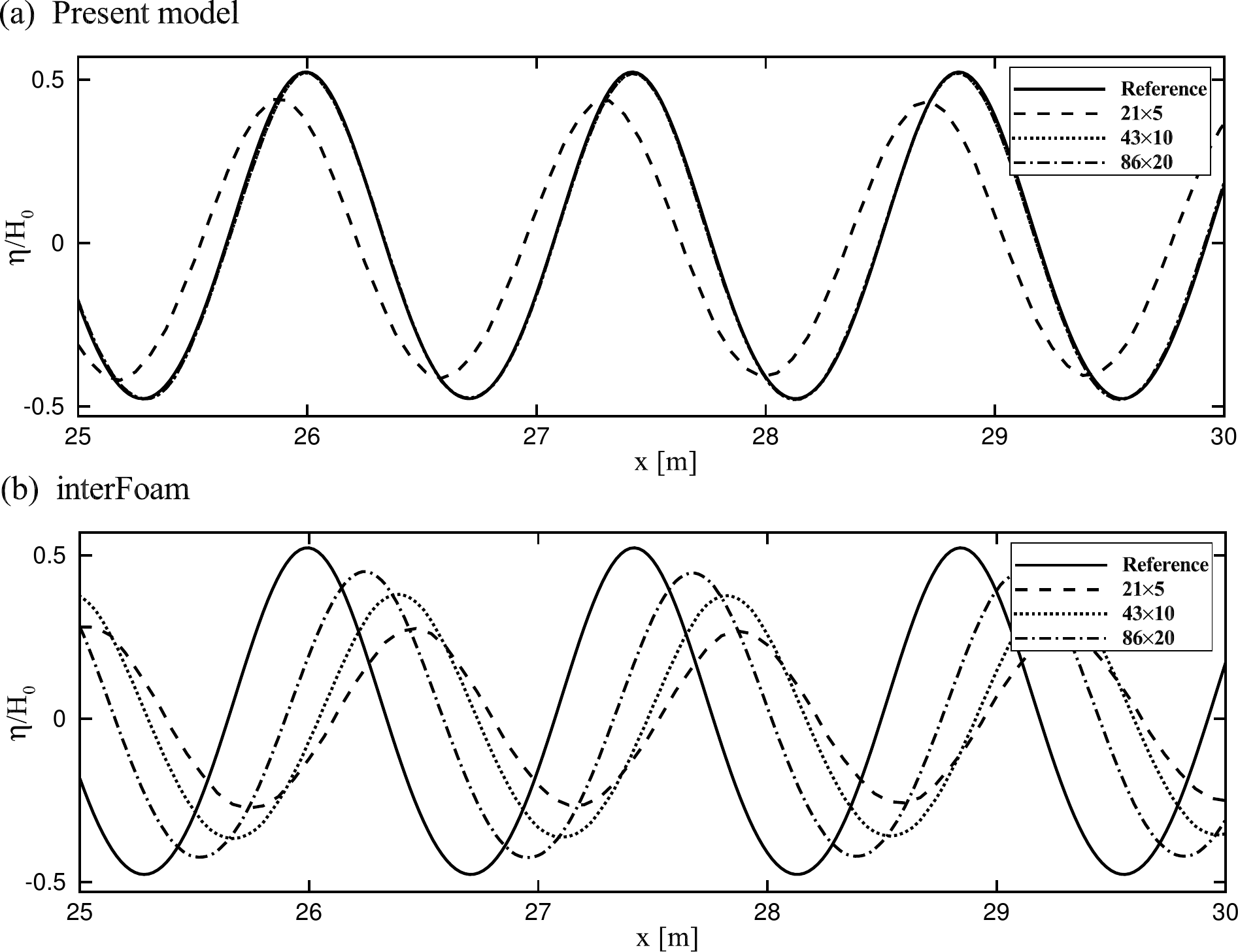}
		\caption{Normalized surface elevations over x-location at $t=60T_{0}$ for different grid resolution.}
	\label{fig:gridAlongx}
	\end{minipage}
\end{figure}

\begin{figure}[!]
	\centering
		\includegraphics[width=0.5\textwidth]{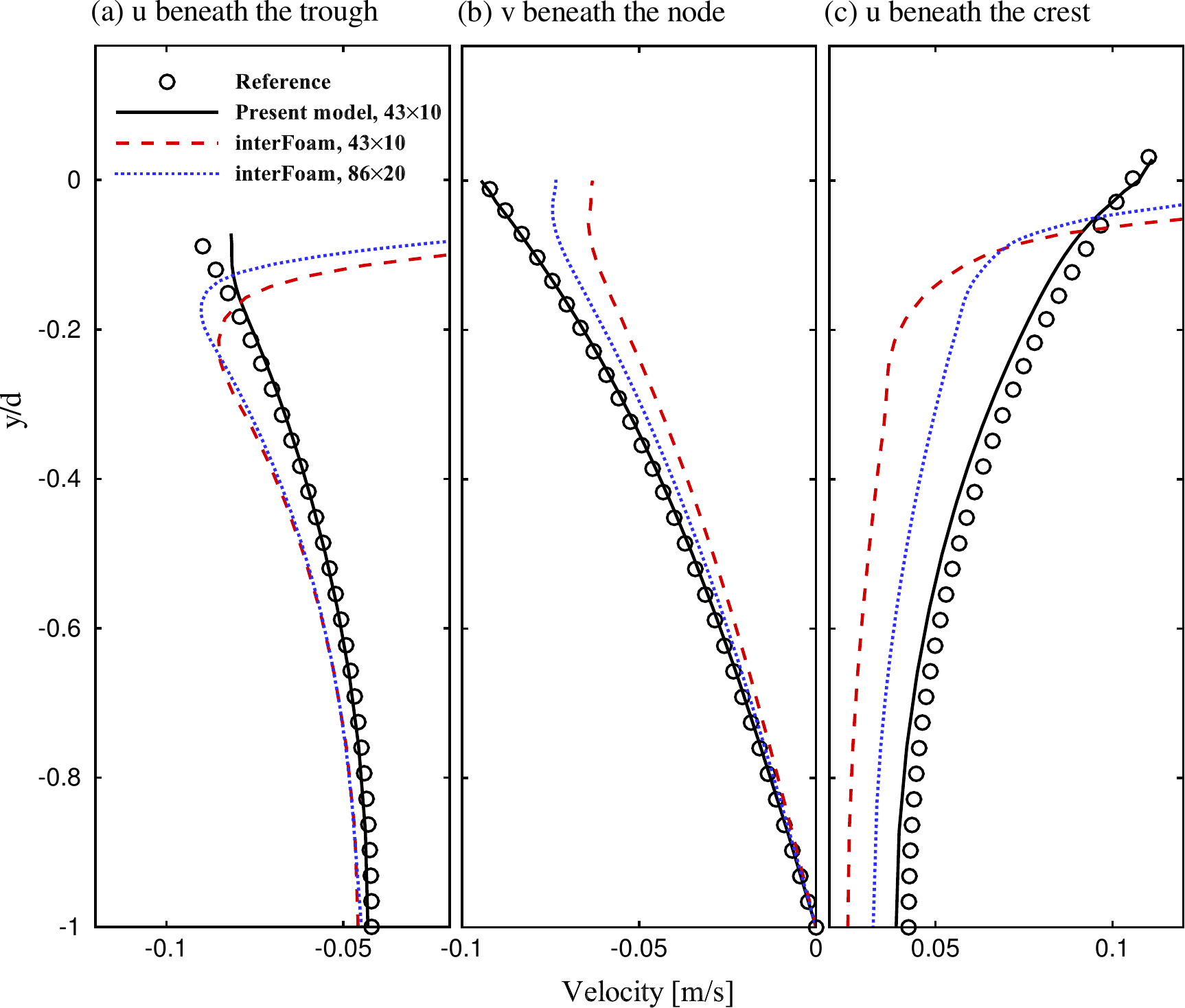}
	\caption{Vertical distribution of velocity components for different grid resolution with both solvers, recorded at different positions of the propagating wave, at $t=60T_{0}$.}
	\label{fig:UVgrid}
\end{figure}

For the grid study, simulations are performed on two other sets of grid, namely, a coarser grid with 21 cells per wavelength and 5 cells per wave height ($21\times5$) and a refined grid with 86 cells per wavelength and 20 cells per wave height ($86\times20$). The time steps are modified at the same time in order to maintain a constant Courant number. As shown in Figs.\ref{fig:gridStudy} and~\ref{fig:gridAlongx}, with the present model, surface elevations are substantially consistent with the reference solutions when relatively fine grids are applied, while a moderate loss in wave height, as well as phase shifts, can be observed with the coarsest grid. The interFoam simulations evidently show that using higher spatial resolution improves the results, however, significant deviations from the reference solutions still can be found even with the finest grid. It is remarkable that the profile phases show opposite features with two numerical solvers. Compared with the reference solutions, the profiles simulated with the present model lag behind in the time series and have a lead over $x$-location, while the interFoam simulations are on the contrary. Following the method utilized in the time step study, we collect the wave parameters in Table~\ref{tab:parameterGrid}. It is demonstrated that the $21\times5$ case of the present model yields a correct wave period but a shorter wavelength compared with the reference solution, where the latter is obviously the cause of the phase errors. Different to the present model, interFoam solver produces shorter wave periods and longer wavelengths, resulting in the phase lag in the time series and phase lead over $x$-location. It is noted that in the $86\times20$ case of the present model, the normalized wave height is a bit over 1.000, the reason for this may include two aspects, the high spatial resolution brings scarcely any numerical dissipation, in the meantime the residual reflection may cause an increase in wave height. Corresponding velocity profiles are shown in Fig.\ref{fig:UVgrid}, similar findings with those in time step study can be noticed from the comparison and will not be elaborated in this section.
\begin{table}[!]
  \small
  \caption{Parameters of simulated waves for the grid study, normalized by the nominal values}
  \centering
    \begin{tabular}{@{\extracolsep{4pt}}lllllll@{}}
    \hline
     \noalign{\smallskip}
     & \multicolumn{3}{l}{Present model} & \multicolumn{3}{l}{interFoam solver}\\
     \noalign{\smallskip}
     \cline{2-4}\cline{5-7}
     \noalign{\smallskip}
     Resolution & $21\times5$ & $43\times10$ & $86\times20$ & $21\times5$ & $43\times10$ & $86\times20$ \\
     \noalign{\smallskip}
     \hline
     \noalign{\smallskip}
     $T/T_{0}$  		   & 1.000  & 1.000  & 1.000  & 0.994  & 0.996  & 0.998 \\
     \noalign{\smallskip}
     $\lambda/\lambda_{0}$ & 0.992  & 0.999  & 0.999  & 1.029  & 1.013  & 1.009 \\
     \noalign{\smallskip}
     $H/H_{0}$  		   & 0.882  & 0.997  & 1.001  & 0.638  & 0.801  & 0.903 \\
     \noalign{\smallskip}
     \hline
    \end{tabular}%
  \label{tab:parameterGrid}%
\end{table}

\subsection{Propagation of waves with different steepness and periods}
After the time step study and grid study, the numerical dissipation in simulating regular waves with various steepness and periods is investigated in this section. The parameters used to define the mass source for different cases are shown in Tables~\ref{tab:steepness} and~\ref{tab:period}, the grid system and the time step of the base case are employed in this section as well, where an identical wave period is applied in the steepness study and vice versa. The maximum wave steepness $H_{0}/\lambda_{0}$ applied in this section is 0.028, which is much smaller than the breaking point of gravity waves, 0.142, this guarantees that no physical dissipation will be introduced into the simulations. Numerical tests are performed with both the present model and interFoam solver. Numerical dissipation is assessed by looking at the wave heights along the wave tank. The same method mentioned in the time step study is utilized to measure the wave heights, in particular, the last 20 periods instead of 30 periods of the time series will be used to compute the value at $x'=20\lambda_{0}$ as it has a relatively late start. 

\begin{figure}[!htb]
	\begin{minipage}[t]{0.48\linewidth}
	\centering
		\includegraphics[width=1.0\textwidth]{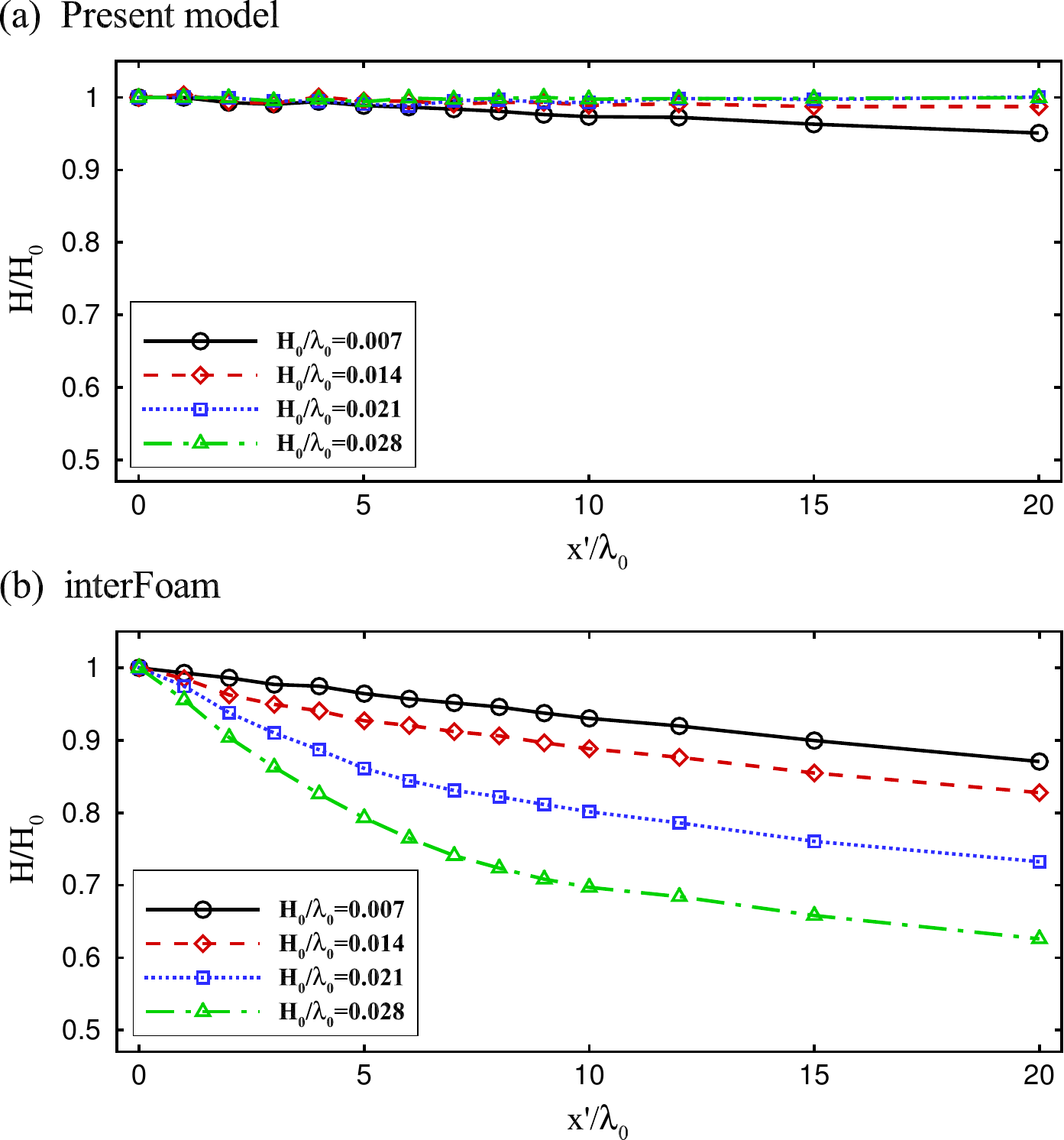}
		\caption{Normalized wave heights along the wave tank for different wave steepness.}
	\label{fig:steepness}
	\end{minipage}
	\hfill
	\begin{minipage}[t]{0.48\linewidth}
	\centering
		\includegraphics[width=1.0\textwidth]{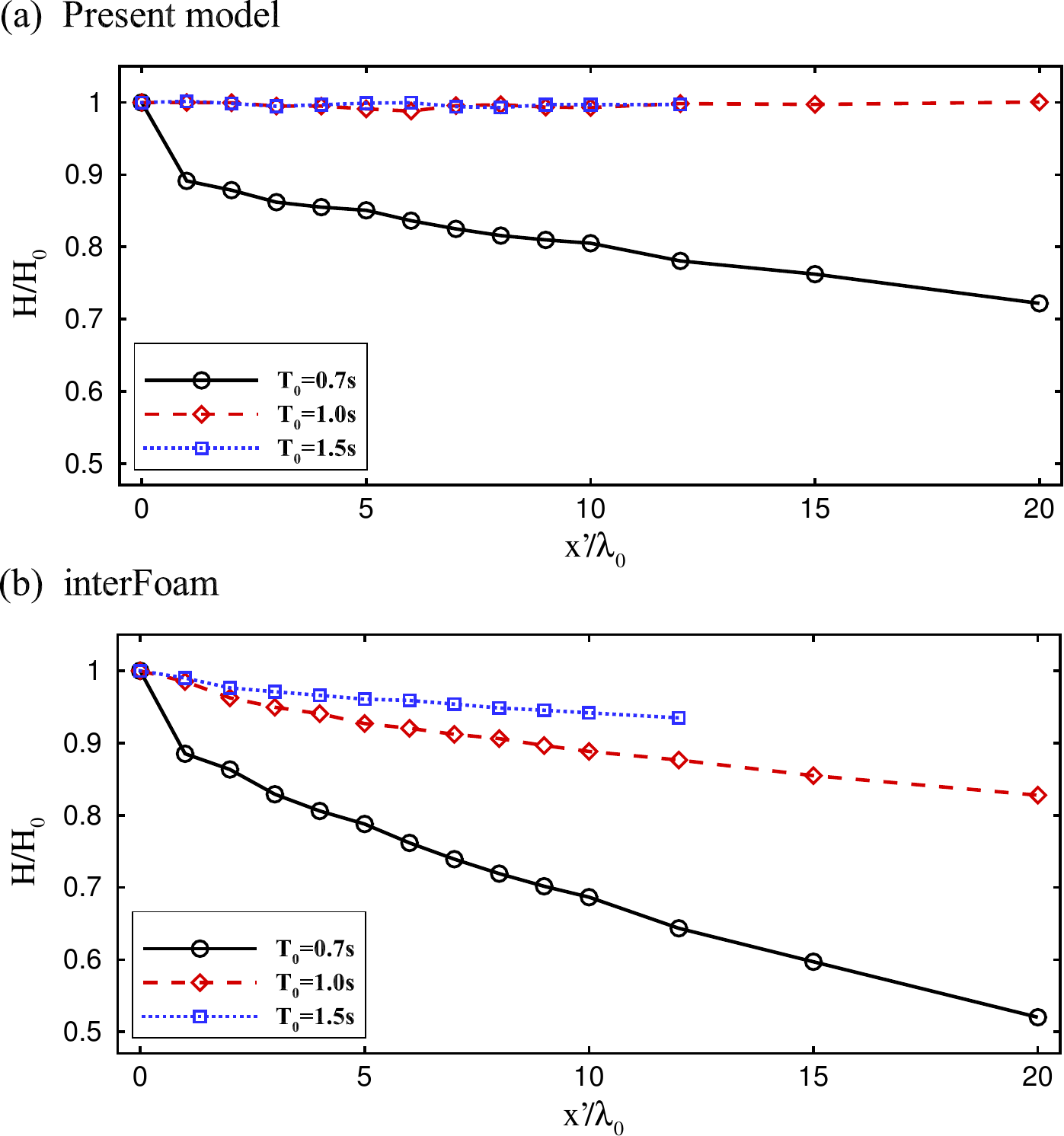}
		\caption{Normalized wave heights along the wave tank for different wave periods.}
	\label{fig:period}
	\end{minipage}
\end{figure}

Wave heights normalized by $H_{0}$ at various distances to the source region are plotted in Figs.\ref{fig:steepness} and~\ref{fig:period}. From Fig.\ref{fig:steepness}, it can be seen that the two numerical solvers have totally different features in propagating waves with various steepness. With the present model, the wave decay remains at a low level, a moderate wave decay is observed in the $H_{0}/\lambda_{0}=0.007$ case, the decay rate decreases as the steepness increases and practically no dissipation can be found in the $H_{0}/\lambda_{0}=0.021, 0.028$ cases. This phenomenon may seem surprising, as one would rather expect an expanding wave decay with increasing wave steepness, which is exactly the feature shown in the interFoam simulations. The wave decay in small steepness cases is believed to lie in the relatively low spatial resolution in $y-$direction since smaller wave heights are adopted (roughly 3 cells per wave height for the $H_{0}/\lambda_{0}=0.007$ case), while for large steepness, the wave decay is mainly attributed to the nonlinearity effects and the large velocity gradients in the vicinity of the free surface. Apparently the present model is efficient in dealing with the large steepness cases but has a limitation in terms of spatial accuracy. Nevertheless, the present model still has a better performance than interFoam solver with regard to small steepness as it only has a maximum wave decay of $5\%$ in the $H_{0}/\lambda_{0}=0.007$ case comparing to that of $13\%$ with interFoam solver. Fig.\ref{fig:period} shows the simulations with various wave periods, the most notable feature is that when simulating a short wave of $T_{0}=0.7$s, both solvers experience a dramatic wave decay, which is $28\%$ for the present model and $48\%$ for interFoam solver at a distance of $20\lambda_{0}$ to the source region. The low spatial resolution in both $x-$ and $y-$directions is supposed to be the primary cause of the massive wave decay, indicating that simulations with small wave periods need to be performed with care even using high-order accurate models, grid refinement is necessary for this sort of simulations. 
\begin{table}[!htb]
  \small
  \caption{Wave parameters used for the simulations with different wave steepness}
  \centering
    \begin{tabular}{@{\extracolsep{4pt}}llll@{}}
     \hline
     \noalign{\smallskip}
     $T_{0}$(s) & $\lambda_{0}$(m) & $H_{0}$(m) & $H_{0}/\lambda_{0}$ \\
     \noalign{\smallskip}
     \hline
     \noalign{\smallskip}
     1.0  & 1.425 & 0.01 & 0.007 \\
     \noalign{\smallskip}
     1.0  & 1.425 & 0.02 & 0.014 \\
     \noalign{\smallskip}
     1.0  & 1.425 & 0.03 & 0.021 \\
     \noalign{\smallskip}
     1.0  & 1.425 & 0.04 & 0.028 \\
     \noalign{\smallskip}
     \hline
    \end{tabular}%
  \label{tab:steepness}%
\end{table}%
\begin{table}[!htb]
  \small
  \caption{Wave parameters used for the simulations with different wave periods}
  \centering
    \begin{tabular}{@{\extracolsep{4pt}}llll@{}}
     \hline
     \noalign{\smallskip}
     $T_{0}$(s) & $\lambda_{0}$(m) & $H_{0}$(m) & $H_{0}/\lambda_{0}$ \\
     \noalign{\smallskip}
     \hline
     \noalign{\smallskip}  
     0.7  & 0.760  & 0.011  & 0.014  \\
     \noalign{\smallskip}
     1.0  & 1.425  & 0.020  & 0.014  \\
     \noalign{\smallskip}
     1.5  & 2.488  & 0.035  & 0.014  \\
     \noalign{\smallskip}
     \hline
    \end{tabular}%
  \label{tab:period}%
\end{table}%
\subsection{Computational cost}
\begin{figure}[H]
	\centering
		\includegraphics[width=0.5\textwidth]{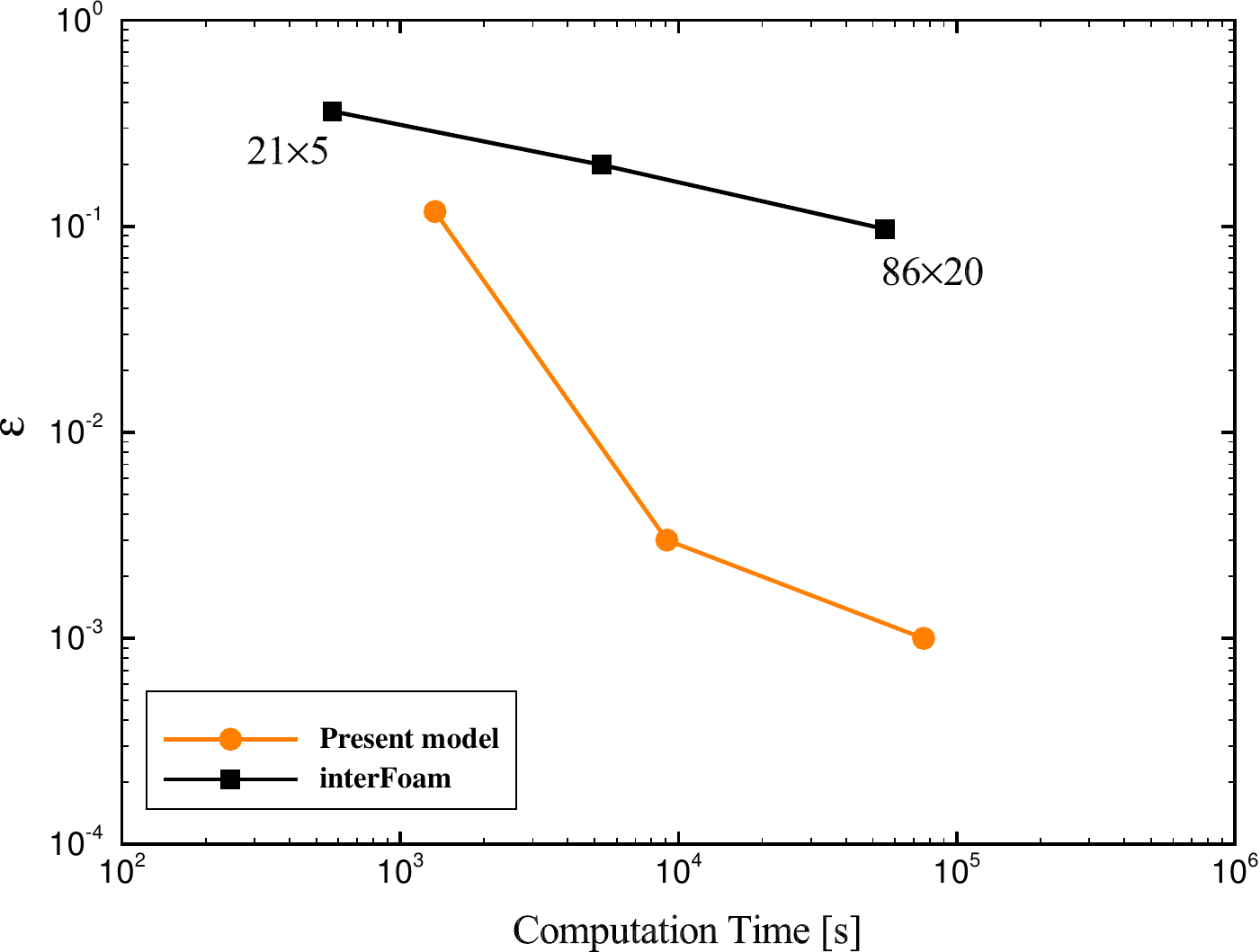}
	\caption{Wave height error against computation time with both solvers.}
	\label{fig:errorGrid}
\end{figure}

The present model has shown particular promise in accuracy benefits, many would be concerned with the computational efficiency
since high-order schemes are usually more complex and computationally expensive compared to the conventional FVM. So in this section, we will quantify the computational cost of both solvers in terms of computation time on equivalent hardware, as well as quantitative accuracy measurements based on suitable error metrics. All the cases from the grid study in section~\ref{GridStudy} are investigated, which run on a processor of Intel Core i7-4790 CPU (3.60GHz) and in each case, only a single core is utilized. To evaluate the accuracy, we consider the wave height error $\varepsilon$ instead of the widely used $L_{2}$ error, where the latter is sensitive to the phase error of the surface elevations that is less concerned in engineering applications. The wave height error $\varepsilon$ is defined as:
\begin{equation}
\label{eq:error}
\varepsilon=\frac{\mid{H-H_{0}\mid}}{H_{0}},
\end{equation}
where the simulated wave height $H$ can be found in Table~\ref{tab:parameterGrid}. A plot of the wave height error against computation time is shown in Fig.\ref{fig:errorGrid}, where the computation time refers to the execution time spent by the processor and is measured in seconds. It can be seen that with both solvers, the computation time increases while the wave height error is reduced as the resolution increases. The present model outperforms interFoam in terms of accuracy on all three sets of grid, while using the present model is more costly than using interFoam for a specific grid system. For the $21\times5$ case, using the present model reduces the error to about $1/3$ compared with interFoam solver, meanwhile it costs $130\%$ more computation time, such a performance-price ratio is remarkable in view of the fact that interFoam solver takes 96 times longer to achieve a comparative accuracy (see the $86\times20$ case of interFoam). In other words, to reach a given error level, for instance, $\varepsilon=\mathcal{O}(10^{-1})$, the present model merely requires a small percentage of computation time that is used by interFoam. Further, as finer grids are applied, the present model demonstrates a faster convergence rate as well as a slower growth in computation time, resulting in an even higher performance-price ratio, with the $86\times20$ case $\approx$2 orders of magnitude more accurate than that of interFoam solver at a cost of 37$\%$ more computation time.
\section{Discussion}
\label{discussion}
\begin{figure}[H]
	\centering
		\includegraphics[width=0.5\textwidth]{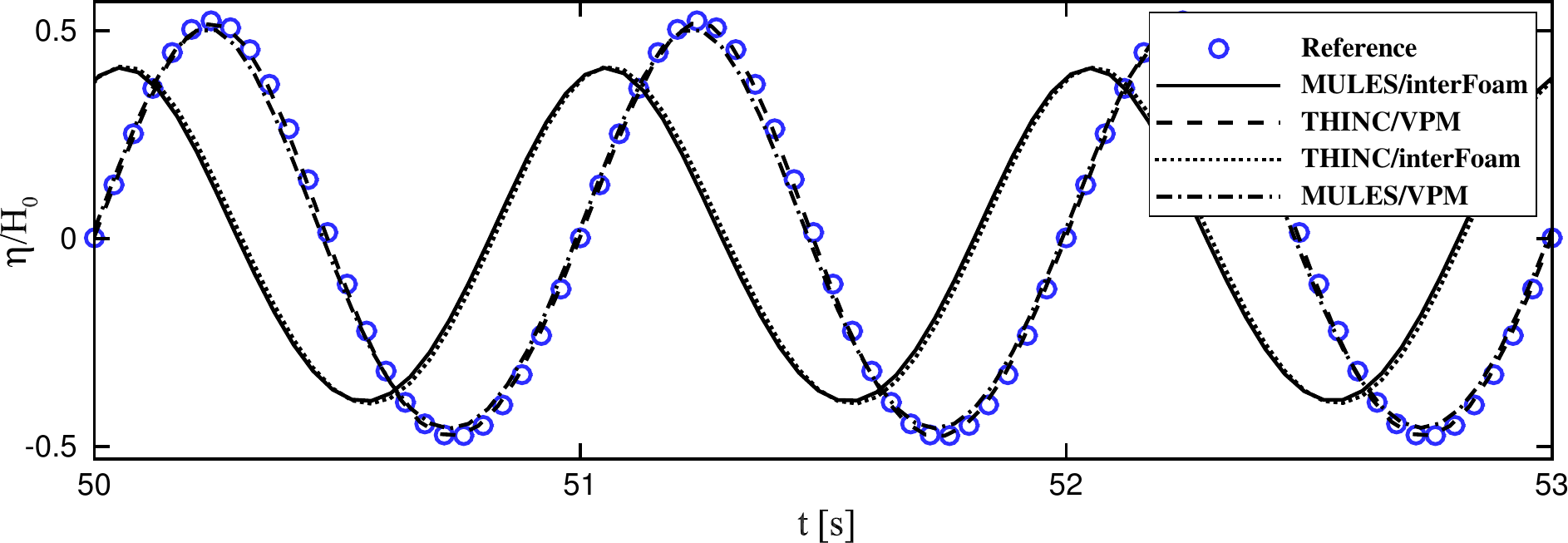}
	\caption{Normalized surface elevations over time with four different models, recorded $10\lambda_{0}$ away from the source region.}
	\label{fig:termsEta}
\end{figure}

\begin{figure}[H]
	\centering
		\includegraphics[width=0.5\textwidth]{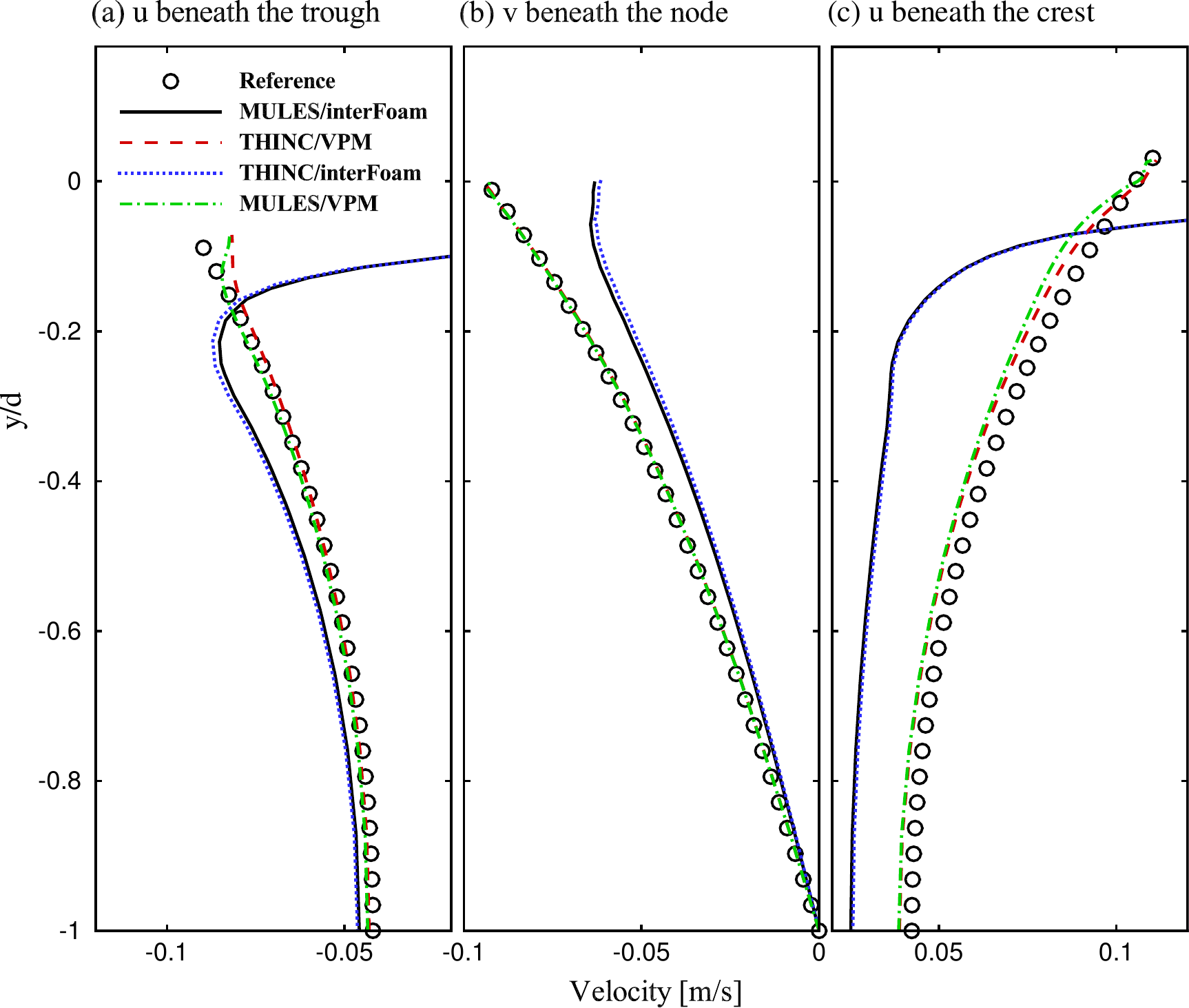}
	\caption{Vertical distribution of velocity components for different grid solutions with four different models, recorded at different positions of the propagating wave, at $t=60T_{0}$.}
	\label{fig:termsUV}
\end{figure}

Although the exact mechanisms of the waveform changes and the velocity oscillations are not clear at present, some numerical tests are carried out to help better understand the issues behind the behavior and narrow down the dominant factors in the solution procedure. Two interface capturing schemes (MULES and THINC/QQ) and two finite volume methods (the second-order accurate FVM applied in interFoam and VPM) compose 4 different solvers and all of them are employed to simulate the base case, corresponding surface elevations and velocity profiles are given in Figs.\ref{fig:termsEta} and~\ref{fig:termsUV}. 
Both the surface elevations and the velocity profiles demonstrate that simulation results with the same momentum solver are close to each other and the VPM-based solvers produce more accurate results, indicating that the impact of the interface capturing scheme is comparatively limited in this sort of applications. In addition, though not shown herein for brevity, we note that the diffusion terms and the surface tension term make little contribution to the behavior as scarcely any difference occurs in surface elevations and velocity fields when they are turned off. Then it is evident that the advection term is the dominant force in this application. Further, we replace the third-order Runge-Kutta scheme with the first-order forward Euler scheme to time advance the advection term, the results vary slightly, suggesting that it is the spatial discretization of the advection term that plays a crucial role in the simulations.

\section{Conclusion remarks}
\label{conclusion}
A 2D wave tank is built in this work, where the VPM scheme and THINC/QQ method are employed to solve the fluid dynamics and capture the free surface, respectively. A mass source function for wave generation and sponge layers for wave absorption are embedded in the present high-order accurate model as well as the open-source code interFoam. The performance of both solvers in simulating the propagation of regular waves is carefully compared in this work.

First, a regular wave in an intermediate depth is simulated as the base case, the surface elevations are compared with the reference solutions given by the second-order Stokes wave theory. It is demonstrated that both wave makers can produce wave trains constantly without significant reflections from the boundary walls. Wave trains are well reproduced inside the tank with the present model and compare well with the reference solutions even after propagating a distance of 20 wavelengths. With interFoam solver, however, simulated waves show a nonnegligible wave decay and phase shifts after traveling 6 wavelengths, which is believed to be attributed to the numerical dissipation. Then a time step study and a grid study are conducted, special attention has been paid to the numerical dissipation along the propagation direction of the wave trains. By examining the surface elevations, numerical dissipation is evaluated in terms of wave heights and phase consistency. The present model shows a great advantage in simulation fidelity as the simulated results are highly consistent with the reference solutions and shown to be basically insensitive to temporal and spatial resolution, only a moderate wave decay and a slight wavelength decrease are found when a relatively coarse grid is applied. The interFoam simulations, on the other hand, show significant wave decay and phase shifts, where the latter are proved to arise from the reduced wave periods and the expanded wavelengths. The results with interFoam solver can be improved by increasing the temporal and spatial resolution, but deviations from the reference solution still exit even using an extremely small time step of 0.0002s that corresponds to a Courant number of 0.01. 

The velocity profiles along the vertical cross-section have also been examined. The present model produces a satisfactory velocity field even with a relatively large time step and a coarse grid. However, the velocities in interFoam simulations are found to be underestimated in the bulk of the water and severely overestimated near the free surface. By reducing the time step to 0.0002s, interFoam solver can have comparative velocity distribution with the present model using a time step of 0.004s. Propagation of regular waves with various steepness and periods is investigated as well. The results indicate that the present model has a much better performance than interFoam solver. While in the short wave case, i.e. $T_{0}=0.7$s, both solvers experience a dramatic wave decay, which is attributed to the relatively low spatial resolution. This is a reminder that simulations involving short-period waves need to be conducted with care, grid refinement is necessary even applying a high-order accurate model. A comparison of the computational cost shows that using the present model causes $37\sim{130}\%$ increase in simulation time for a given grid system, whereas the present model outperforms interFoam solver in terms of accuracy by up to 2 orders of magnitude. Consequently, the present model requires much less computational cost to reach a given error level in comparison with interFoam solver. 

Further model analysis suggests that the spatial discretization of the advection term, instead of the interface capturing scheme as many would expect, plays a crucial role in this application. Although the interface capturing scheme THINC/QQ has not demonstrated its advantages over MULES scheme in simulating the propagation of regular waves, it is believed to be promising for applications involving violently-changing interfaces, for instance, wave breaking and wave-structure interactions. 

\section*{Acknowledgements}
The early stages of this study were conducted during the first author's stay at Tokyo Institute of Technology. The authors are grateful to the helpful discussions with some members of Xiao Lab. Special acknowledgements should be given to Dr. Feng Xiao. This study was partially supported by the National Natural Science Foundation of China (Grant Nos. 51479175, 51679212), Zhejiang Provincial Natural Science Foundation of China (Grant No. R16E090002).

\bibliographystyle{IEEEtran}
\bibliography{myRef} 

\begin{thebibliography}{10}
\providecommand{\url}[1]{#1}
\csname url@samestyle\endcsname
\providecommand{\newblock}{\relax}
\providecommand{\bibinfo}[2]{#2}
\providecommand{\BIBentrySTDinterwordspacing}{\spaceskip=0pt\relax}
\providecommand{\BIBentryALTinterwordstretchfactor}{4}
\providecommand{\BIBentryALTinterwordspacing}{\spaceskip=\fontdimen2\font plus
\BIBentryALTinterwordstretchfactor\fontdimen3\font minus
  \fontdimen4\font\relax}
\providecommand{\BIBforeignlanguage}[2]{{%
\expandafter\ifx\csname l@#1\endcsname\relax
\typeout{** WARNING: IEEEtran.bst: No hyphenation pattern has been}%
\typeout{** loaded for the language `#1'. Using the pattern for}%
\typeout{** the default language instead.}%
\else
\language=\csname l@#1\endcsname
\fi
#2}}
\providecommand{\BIBdecl}{\relax}
\BIBdecl

\bibitem{jacobsen2012wave}
N.~G. Jacobsen, D.~R. Fuhrman, and J.~Freds{\o}e, ``A wave generation toolbox
  for the open-source cfd library: Openfoam{\textregistered},''
  \emph{International Journal for Numerical Methods in Fluids}, vol.~70, no.~9,
  pp. 1073--1088, 2012.

\bibitem{higuera2013realistic}
P.~Higuera, J.~L. Lara, and I.~J. Losada, ``Realistic wave generation and
  active wave absorption for navier--stokes models: Application to
  openfoam{\textregistered},'' \emph{Coastal Engineering}, vol.~71, pp.
  102--118, 2013.

\bibitem{roenby2017new}
J.~Roenby, B.~E. Larsen, H.~Bredmose, and H.~Jasak, ``A new volume-of-fluid
  method in openfoam,'' in \emph{VII International Conference on Computational
  Methods in Marine Engineering, MARINE}, vol. 2017, 2017, pp. 1--12.

\bibitem{jacobsen2014formation}
N.~G. Jacobsen, J.~Fredsoe, and J.~H. Jensen, ``Formation and development of a
  breaker bar under regular waves. part 1: Model description and
  hydrodynamics,'' \emph{Coastal Engineering}, vol.~88, pp. 182--193, 2014.

\bibitem{jacobsen2014formation02}
N.~G. Jacobsen and J.~Fredsoe, ``Formation and development of a breaker bar
  under regular waves. part 2: Sediment transport and morphology,''
  \emph{Coastal Engineering}, vol.~88, pp. 55--68, 2014.

\bibitem{brown2016evaluation}
S.~Brown, D.~Greaves, V.~Magar, and D.~Conley, ``Evaluation of turbulence
  closure models under spilling and plunging breakers in the surf zone,''
  \emph{Coastal Engineering}, vol. 114, pp. 177--193, 2016.

\bibitem{higuera2013simulating}
P.~Higuera, J.~L. Lara, and I.~J. Losada, ``Simulating coastal engineering
  processes with openfoam{\textregistered},'' \emph{Coastal Engineering},
  vol.~71, pp. 119--134, 2013.

\bibitem{peric2015generation}
R.~Peri{\'c} and M.~Abdel-Maksoud, ``Generation of free-surface waves by
  localized source terms in the continuity equation,'' \emph{Ocean
  Engineering}, vol. 109, pp. 567--579, 2015.

\bibitem{cha2011numerical}
J.-J. Cha and D.-C. Wan, ``Numerical wave generation and absorption based on
  openfoam,'' \emph{Ocean Engineering(Haiyang Gongcheng)}, vol.~29, no.~3, pp.
  1--12, 2011.

\bibitem{larsen2018performance}
B.~E. Larsen, D.~R. Fuhrman, and J.~Roenby, ``Performance of interfoam on the
  simulation of progressive waves,'' \emph{arXiv preprint arXiv:1804.01158},
  2018.

\bibitem{wroniszewski2014benchmarking}
P.~A. Wroniszewski, J.~C. Verschaeve, and G.~K. Pedersen, ``Benchmarking of
  navier--stokes codes for free surface simulations by means of a solitary
  wave,'' \emph{Coastal Engineering}, vol.~91, pp. 1--17, 2014.

\bibitem{jin2014numerical}
H.~Jin, Y.~Liu, S.-y. He, H.-j. Li \emph{et~al.}, ``Numerical study on the wave
  dissipating performance of a submerged horizontal plate breakwater using
  openfoam,'' in \emph{The Eleventh ISOPE Pacific/Asia Offshore Mechanics
  Symposium}.\hskip 1em plus 0.5em minus 0.4em\relax International Society of
  Offshore and Polar Engineers, 2014.

\bibitem{hu2016numerical}
Z.~Z. Hu, D.~Greaves, and A.~Raby, ``Numerical wave tank study of extreme waves
  and wave-structure interaction using openfoam{\textregistered},'' \emph{Ocean
  Engineering}, vol. 126, pp. 329--342, 2016.

\bibitem{windt2017assessment}
C.~Windt, J.~Davidson, P.~Schmitt, and J.~V. Ringwood, ``Assessment of
  numerical wave makers,'' in \emph{Proceedings of the 12th European Wave and
  Tidal Energy Conference}, 2017.

\bibitem{cockburn1989tvb}
B.~Cockburn and C.-W. Shu, ``Tvb runge-kutta local projection discontinuous
  galerkin finite element method for conservation laws. ii. general
  framework,'' \emph{Mathematics of computation}, vol.~52, no. 186, pp.
  411--435, 1989.

\bibitem{wang2002spectral}
Z.~J. Wang, ``Spectral (finite) volume method for conservation laws on
  unstructured grids. basic formulation: Basic formulation,'' \emph{Journal of
  computational physics}, vol. 178, no.~1, pp. 210--251, 2002.

\bibitem{ii20054th}
S.~Ii, M.~Shimuta, and F.~Xiao, ``A 4th-order and single-cell-based advection
  scheme on unstructured grids using multi-moments,'' \emph{Computer physics
  communications}, vol. 173, no. 1-2, pp. 17--33, 2005.

\bibitem{xiao2006unified}
F.~Xiao, R.~Akoh, and S.~Ii, ``Unified formulation for compressible and
  incompressible flows by using multi-integrated moments ii: Multi-dimensional
  version for compressible and incompressible flows,'' \emph{Journal of
  Computational Physics}, vol. 213, no.~1, pp. 31--56, 2006.

\bibitem{chen2008shallow}
C.~Chen and F.~Xiao, ``Shallow water model on cubed-sphere by multi-moment
  finite volume method,'' \emph{Journal of Computational Physics}, vol. 227,
  no.~10, pp. 5019--5044, 2008.

\bibitem{xie2014multi}
B.~Xie, S.~Ii, A.~Ikebata, and F.~Xiao, ``A multi-moment finite volume method
  for incompressible navier--stokes equations on unstructured grids:
  volume-average/point-value formulation,'' \emph{Journal of Computational
  Physics}, vol. 277, pp. 138--162, 2014.

\bibitem{xie2017toward}
B.~Xie and F.~Xiao, ``Toward efficient and accurate interface capturing on
  arbitrary hybrid unstructured grids: The thinc method with quadratic surface
  representation and gaussian quadrature,'' \emph{Journal of Computational
  Physics}, vol. 349, pp. 415--440, 2017.

\bibitem{xie2017unstructured}
B.~Xie, P.~Jin, and F.~Xiao, ``An unstructured-grid numerical model for
  interfacial multiphase fluids based on multi-moment finite volume formulation
  and thinc method,'' \emph{International Journal of Multiphase Flow}, vol.~89,
  pp. 375--398, 2017.

\bibitem{lin1999internal}
P.~Lin and P.~L.-F. Liu, ``Internal wave-maker for navier-stokes equations
  models,'' \emph{Journal of waterway, port, coastal, and ocean engineering},
  vol. 125, no.~4, pp. 207--215, 1999.

\bibitem{wei1995time}
G.~Wei and J.~T. Kirby, ``Time-dependent numerical code for extended boussinesq
  equations,'' \emph{Journal of Waterway, Port, Coastal, and Ocean
  Engineering}, vol. 121, no.~5, pp. 251--261, 1995.

\bibitem{prosperetti2009computational}
A.~Prosperetti and G.~Tryggvason, \emph{Computational methods for multiphase
  flow}.\hskip 1em plus 0.5em minus 0.4em\relax Cambridge university press,
  2009.

\bibitem{chorin1968numerical}
A.~J. Chorin, ``Numerical solution of the navier-stokes equations,''
  \emph{Mathematics of computation}, vol.~22, no. 104, pp. 745--762, 1968.

\bibitem{gottlieb1998total}
S.~Gottlieb and C.-W. Shu, ``Total variation diminishing runge-kutta schemes,''
  \emph{Mathematics of computation of the American Mathematical Society},
  vol.~67, no. 221, pp. 73--85, 1998.

\bibitem{jasak1996error}
H.~Jasak, ``Error analysis and estimation for the finite volume method with
  applications to fluid flows.'' 1996.

\bibitem{deshpande2012evaluating}
S.~S. Deshpande, L.~Anumolu, and M.~F. Trujillo, ``Evaluating the performance
  of the two-phase flow solver interfoam,'' \emph{Computational science \&
  discovery}, vol.~5, no.~1, p. 014016, 2012.

\bibitem{paulsen2014forcing}
B.~T. Paulsen, H.~Bredmose, H.~B. Bingham, and N.~G. Jacobsen, ``Forcing of a
  bottom-mounted circular cylinder by steep regular water waves at finite
  depth,'' \emph{Journal of fluid mechanics}, vol. 755, pp. 1--34, 2014.

\bibitem{francois2006balanced}
M.~M. Francois, S.~J. Cummins, E.~D. Dendy, D.~B. Kothe, J.~M. Sicilian, and
  M.~W. Williams, ``A balanced-force algorithm for continuous and sharp
  interfacial surface tension models within a volume tracking framework,''
  \emph{Journal of Computational Physics}, vol. 213, no.~1, pp. 141--173, 2006.

\bibitem{wemmenhove2015numerical}
R.~Wemmenhove, R.~Luppes, A.~E. Veldman, and T.~Bunnik, ``Numerical simulation
  of hydrodynamic wave loading by a compressible two-phase flow method,''
  \emph{Computers \& Fluids}, vol. 114, pp. 218--231, 2015.

\end{thebibliography}

\end{document}